\documentclass[aps, prd, reprint, nofootinbib, longbibliography]{revtex4-1}

\usepackage{graphics,graphicx,epsfig}
\usepackage{amsmath, amssymb, color}
\usepackage[dvipsnames]{xcolor}
\usepackage[colorlinks=true, linkcolor=blue, citecolor=magenta]{hyperref}
\usepackage{url}

\usepackage{float, placeins}
\usepackage{mwe}
\usepackage[caption=false]{subfig}
\usepackage{morefloats}
\usepackage{empheq}
\usepackage{bm}
\usepackage{booktabs} 
\usepackage{multirow}
\usepackage{makecell}

\newcommand\Mpl{M_{\textrm{pl}}}

\begin{document}

\title{Gauging Fine-Tuning}

\author{Feraz Azhar}
\email[Email address: {feraz\_azhar@fas.harvard.edu}]{}
\affiliation{Black Hole Initiative, Harvard University, Cambridge, Massachusetts 02138, USA}

\author{Abraham Loeb}
\email[Email address: aloeb@cfa.harvard.edu]{}
\affiliation{Black Hole Initiative, Harvard University, Cambridge, Massachusetts 02138, USA}

\date{\today}

\begin{abstract}
We introduce a mathematical framework for quantifying fine-tuning in general physical settings. In particular, we identify two distinct perspectives on fine-tuning, namely, a local and a global perspective --- and develop corresponding measures. These measures apply broadly to settings characterized by an arbitrary number of observables whose values are dependent on an arbitrary number of parameters. We illustrate our formalism by quantifying fine-tuning as it arises in two pertinent astrophysical settings: (i) in models where a significant fraction of the dark matter in the universe is in the form of primordial black holes, and (ii) in scenarios that derive the fraction of protons in habitable dark-matter halos from underlying models of cosmic inflation.
 
\end{abstract}

\maketitle

\section{Introduction\label{SEC:Introduction}}

The fine-tuning of our existence, as encoded in our current best (effective) physical theories --- such as the standard models of particle physics and cosmology --- is a striking putative fact~\citep{carter_74, carr+rees_79, barrow+tipler_86}. And yet this is just one instance of fine-tuning as it arises for such theories. More broadly, we may identify the fine-tuning of some particular phenomenon, $\mathcal{F}$, in the context of some theory $\mathcal{T}$, in the following way (for which we provide a mnemonic label, as the claim will recur below):
\begin{quote}
(FT): {\it If circumstances in theory $\mathcal{T}$ were a little different, phenomenon $\mathcal{F}$ would not arise.}
\end{quote}
Although simple to state, there are significant technical and conceptual challenges involved in making this claim precise.

The `circumstances' referred to in (FT), are perhaps less controversial. They refer, in practice, to three items that characterize an effective physical theory: (i) equations of evolution for dynamical variables of interest, (ii) initial conditions for those dynamical variables, and (iii) parameters that appear in the equations. These parameters are left unspecified by any such theory, and are fixed by our observations. For example,~\citet{tegmark+al_06} identify a total of 31 such parameters that arise in the standard models of particle physics (26 parameters) and cosmology (5 parameters).

A particularly difficult challenge involves making precise the notion of `a little different'. This problem sits at the heart of what we mean by fine-tuning, and quantifying this notion is the focus of this paper. 

One also needs to be clear about which phenomenon, $\mathcal{F}$, one is referring to in (FT). There are cases where $\mathcal{F}$ is straightforward to identify, but where any subsequent analysis would be uninteresting. For example, ordinary objects on Earth appear to be finely tuned, but their existence can be accounted for by a sequence of accidents --- in which case their finely tuned nature does not seem to require an explanation. Thus, the question arises: what are the salient phenomena (the appropriate $\mathcal{F}$'s) that should be the target of our concerns about fine-tuning? As mentioned in the opening paragraph, this question is of particular interest (and is particularly difficult to answer) when we would like $\mathcal{F}$ to stand for `our existence'. (See Ref.~\cite{maudlin_crete} for a discussion of this point.)

A second facet of the issue of fine-tuning, aside from concerns that relate to its definition, arises when we ask: how should we deal with theories that predict phenomena that are finely tuned? For now, we will refer to such theories as `finely tuned theories'. One can think of finely tuned theories as containing `finely tuned models', which instantiate the theory in the real world --- though we will not make much of this distinction in what follows. One response, that we broadly endorse, is that finely tuned theories seem to cry out for a replacement, namely, a less-finely tuned theory, where phenomena of interest do not disappear under small changes in `circumstances'. And there are a variety of examples in physics of new, putatively less finely tuned theories supplanting more finely tuned theories. An interesting example from the recent history of physics is that of cosmic inflation (see Refs.~\citep{guth_81, linde_82, albrecht+steinhardt_82, linde_83b}) supplanting the standard big-bang model of cosmology.\footnote{And, of course, one can wonder whether cosmic inflation is itself finely tuned --- see, for a lively recent debate, Refs.~\citep{ijjas+al_13, guth+al_14}.} (We touch upon further examples, gleaned more broadly from the history of science, in Sec.~\ref{SEC:Discussion}.)

In this paper, motivated by (a) technical challenges inherent in making claims about fine-tuning precise, and (b) the role that fine-tuning plays in the development of theories, we `gauge fine-tuning' --- namely, we develop quantitative measures of fine-tuning suited to the analysis of various models employed in the sciences. (See Refs.~\cite{ellis+al_86, barbieri+giudice_88, anderson+castano_95, athron+miller_07} for related work in the context of particle physics.) Our goal is to provide a means to compare levels of fine-tuning between models, to aid in the task of theory evaluation and development. We define both local and global measures of fine-tuning, and then illustrate our formalism by applying these measures to quantify levels of fine-tuning in two pertinent astrophysical examples. In particular, we (i) characterize fine-tuning in models where a significant fraction of the dark matter in the universe is in the form of primordial black holes (PBHs), and (ii) characterize fine-tuning in more realistic models that derive the fraction of protons in habitable dark-matter halos from underlying models of cosmic inflation. 

In Sec.~\ref{SEC:Measures} we describe the general setting in which our analysis of fine-tuning most naturally applies, and then define local and global measures of fine-tuning. In Sec.~\ref{SEC:PBHs} we apply these measures to characterize fine-tuning in models of dark matter based on PBHs, and then, in Sec.~\ref{SEC:HabitableHalos}, characterize fine-tuning in models of habitable dark-matter halos derived from cosmic inflation. We summarize our conclusions in Sec.~\ref{SEC:Discussion}.

\section{Measures of fine-tuning}\label{SEC:Measures}

We begin by outlining the general setting in which we will define local and global measures of fine-tuning. Consider a theory (or a model) in which there is a total of $m$ observables, represented by a vector $\vec{O}\in \mathbb{R}^{m}$, which arise via a map from a set of $n$ parameters, represented by a vector $\bm{p}\in\mathcal{P}\subset \mathbb{R}^{n}$, with
\begin{align}
\vec{O}: \mathcal{P} &\longrightarrow \mathbb{R}^{m}\nonumber \\
   \bm{p} &\longmapsto \vec{O}(\bm{p}) \equiv \left(O_{1}(\bm{p}),O_{2}(\bm{p}),\dots,O_{m}(\bm{p})\right).\label{EQN:ObservableMapping}
\end{align}
We assume that each parameter $p_{i}$, for $i=1,\dots,n$, is a real number that takes values in some \emph{finite} interval $\mathcal{P}_{i}\subset\mathbb{R}$, and that `parameter space', denoted by $\mathcal{P}=\mathcal{P}_{1}\times \mathcal{P}_{2}\times\dots\mathcal{P}_{n}$ (viz., the $n$-ary Cartesian product of such finite intervals), is  an $n$-dimensional box in $\mathbb{R}^n$. (One can think of $\vec{O}$ as a collection of salient phenomena --- namely, a collection of $\mathcal{F}$s --- as introduced in Sec.~\ref{SEC:Introduction}.) 

The justification for restricting attention to such a finite parameter space is subtle (see, e.g., Ref.~\cite{barnes_17} for a related discussion). As mentioned in Sec.~\ref{SEC:Introduction}, in practice, our characterizations of fine-tuning occur for effective physical theories, where those theories have a limited regime of applicability. For example, the standard models of particle physics and cosmology do not apply to energy scales where effects of quantum gravity become important. This limit on a maximum energy scale translates into limits on dimensionful parameters in the theory (a minimum energy scale can be reasonably assumed to be zero). When dimensionless parameters in the theory determine dimensionful physical quantities (what~\citet{tegmark+al_06} refer to as ``derived physical parameters''), such as masses of particles, then they too may be reasonably assumed to have finite ranges --- so that the resulting physical quantities do not lie outside the bounds of applicability of the effective theory. And when one is faced with a parameter whose range cannot be otherwise restricted (or where one side of the range cannot be restricted), one must use background knowledge to restrict the range in order to capture salient features of the theory. In sum: for each parameter $p_{i}$, the interval $\mathcal{P}_{i}$ is set either by the regime of applicability of the effective theory under investigation, or else by our expectations about values of $p_{i}$ that characterize the theory. (We will see how such considerations enter into our characterization of fine-tuning in Secs.~\ref{SEC:PBHs} and \ref{SEC:HabitableHalos}.)

We note one further feature of the general setting in which we examine issues of fine-tuning. The map from parameters to observables, viz., Eq.~(\ref{EQN:ObservableMapping}), is designed to be general, in that it can include rather complicated conjunctions of dynamical evolutions and algebraic maps that may take us from the parameters of the theory (as encoded in $\bm{p}$), to the measurements we make in our experiments (as encoded in $\vec{O}$). Such a feature will be at play in our examples in Secs.~\ref{SEC:PBHs} and \ref{SEC:HabitableHalos}. 

\subsection{Local measure of fine-tuning}\label{SEC:LocalMeasure}

We denote the local sensitivity of the $\mu$th observable $O_{\mu}$, at $\bm{p'}\in\mathbb{R}^{n}$, in some direction in parameter space $\hat{\bm{\epsilon}}\equiv\bm{\epsilon}/\vert\bm{\epsilon}\vert$, by $\mathcal{L}_{\mu}(\bm{p'};\hat{\bm{\epsilon}})$, and define this via a dimensionless, fractional change in $O_{\mu}$:
\begin{subequations}
\begin{align}
\mathcal{L}_{\mu}(\bm{p'}; \hat{\bm{\epsilon}}) & \equiv \lim_{|\bm{\epsilon}|\to 0}
\frac{\left[O_{\mu}(\bm{p'}+\bm{\epsilon})-O_{\mu}(\bm{p'})\right]/|O_{\mu}(\bm{p'})|}{|\bm{\epsilon}|/|\bm{p'}|}\label{EQN:Localfirst}\\
& = \frac{|\bm{p'}|}{|O_{\mu}(\bm{p'})|}\sum_{i=1}^{n}  \left.\frac{\partial O_{\mu}}{\partial p_{i}}\right\vert_{\bm{p'}}\frac{\epsilon_{i}}{|\bm{\epsilon}|} \label{EQN:Localsecond}\\
&= \frac{|\bm{p'}|}{|O_{\mu}(\bm{p'})|}\left(\hat{\bm{\epsilon}}\cdot\bm{\nabla}O_{\mu}\right)\vert_{\bm{p'}}.
\end{align}
\end{subequations}
We note that the limit in Eq.~(\ref{EQN:Localfirst}) signals that we take the term that is of lowest order in $\vert\bm{\epsilon}\vert$, as displayed in Eq.~(\ref{EQN:Localsecond}). When $O_{\mu}(\bm{p'})>0$, we find that
\begin{equation}\label{EQN:LocalSimp}
\mathcal{L}_{\mu}(\bm{p'}; \hat{\bm{\epsilon}}) = |\bm{p'}|\left(\hat{\bm{\epsilon}}\cdot\bm{\nabla}\ln O_{\mu}\right)\vert_{\bm{p'}},
\end{equation}
namely, at any point in parameter space $\bm{p'}$, the local sensitivity of the $\mu$th observable is proportional to the projection of the gradient of the logarithm of the $\mu$th observable onto the direction of interest in parameter space (viz., $\hat{\bm{\epsilon}}$).

To construct a general measure of the degree of local fine-tuning taking into account all the observables, which we will denote by $\mathcal{L}(\bm{p'}; \hat{\bm{\epsilon}})$, we combine local sensitivities of each individual observable via the usual Euclidean norm:
\begin{align}
\mathcal{L}(\bm{p'}; \hat{\bm{\epsilon}}) &\equiv \left\{\sum_{\mu=1}^{m}\left[\mathcal{L}_{\mu}(\bm{p'}; \hat{\bm{\epsilon}})\right]^{2}\right\}^{1/2} \nonumber\\
&= \left\{\sum_{\mu=1}^{m}\left[\frac{|\bm{p'}|}{|O_{\mu}(\bm{p'})|}\left(\hat{\bm{\epsilon}}\cdot\bm{\nabla}O_{\mu}\right)\vert_{\bm{p'}}
\right]^{2}\right\}^{1/2}.\label{EQN:LocalCom}
\end{align}
If we have just a single observable ($m=1$), as will be the case for the examples considered in Secs.~\ref{SEC:PBHs} and~\ref{SEC:HabitableHalos}, setting $O_{1}\to O$, we find
\begin{equation}
\mathcal{L}(\bm{p'}; \hat{\bm{\epsilon}}) \equiv |\mathcal{L}_{1}(\bm{p'}; \hat{\bm{\epsilon}})|= \left\vert\frac{|\bm{p'}|}{|O(\bm{p'})|}\left(\hat{\bm{\epsilon}}\cdot\bm{\nabla}O\right)\vert_{\bm{p'}}
\right\vert.\label{EQN:Lform1}
\end{equation}

We distinguish orders of local fine-tuning based on the value of $\mathcal{L}$: 
\begin{subequations}\label{EQN:LocalClassif}
\begin{align}
\mathcal{L}(\bm{p'}; \hat{\bm{\epsilon}}) = 0 &\equiv\textrm{Order }0,\\
0<\mathcal{L}(\bm{p'}; \hat{\bm{\epsilon}})\leqslant 1 &\equiv\textrm{Order }1,\\
1<\mathcal{L}(\bm{p'}; \hat{\bm{\epsilon}})\leqslant 2 &\equiv\textrm{Order }2,\\
&\vdots\nonumber\\
N-1<\mathcal{L}(\bm{p'}; \hat{\bm{\epsilon}})\leqslant N &\equiv\textrm{Order }N,
\end{align}
\end{subequations}
where $N\in\{1,2,\dots\}$ can, in principle, be unbounded. When $\mathcal{L}=0$, there is no local fine-tuning. Though, depending on the scenario at issue, values of $\mathcal{L}$ very close to zero may not, in fact, be significantly different from zero.

A motivation for such a classification scheme comes from noting that in particular instances, $\mathcal{L}$ is the exponent in a power-law dependence of observables on parameters, viz., $O\sim p^{\pm\mathcal{L}}$. [One can explicitly derive this relation from, for example, Eq.~(\ref{EQN:Lform1}): wherein we fix the total number of parameters to be one (i.e., $n=1$); set that parameter (denoted by $p$) to be positive ($p>0$); and choose $O(p) > 0$.] In such cases, values of $\mathcal{L}$ that correspond to different orders of local fine-tuning are, indeed, significantly different from each other. (Of course, such values of $\mathcal{L}$ would be even more significantly different from each other if we had characterized the dependence of observables on parameters via an exponential relationship.)

Now, local measures of fine-tuning do not provide an exhaustive account of the finely tuned nature of some observable (or set of observables). For if $\hat{\bm{\epsilon}}$ points along a contour of, say, $O(\bm{p'})$, then $\mathcal{L}(\bm{p'};\hat{\bm{\epsilon}}) = 0$. But we do not expect this fact to be sufficient to definitively rule out fine-tuning of the observable. What is also needed is a global measure of fine-tuning, and we turn to such a measure in the following section. 

\subsection{Global measure of fine-tuning}\label{SEC:GlobalMeasure}

Our global measure of fine-tuning accords most closely with the common notion of a finely tuned observable [as implicit in (FT) in Sec.~\ref{SEC:Introduction}]. That is, we commonly think of an observable as being finely tuned with respect to some parameter if a small, finite change in that parameter (i.e., not an infinitesimal change), leads to a large change in the value of the observable. By `a small finite change' of a parameter, we usually mean some (measurably) small fraction of the range of values that the parameter could have taken. Here, we will develop the case where that measure is the usual Euclidean norm (but our scheme is consistent with more complicated measures as well). Thus, put in plain terms, our global measure of fine-tuning compares (i) how far one needs to move in some direction in parameter space for a significant change in some observable (or set of observables) to occur, with (ii) how far one could have traveled in principle, in that direction in parameter space. (Recall that our parameter space, $\mathcal{P}$, is finite in all directions.)

The construction of our global measure of fine-tuning proceeds as follows. At some point in parameter space $\bm{p'}$, in some direction in parameter space $\hat{\bm{v}}$, we find the length of the vector, $\vert{\bm{v}}\vert$, such that one obtains an order-unity fractional change in the observable (i.e., a large change in the observable). So we need to find $\vert{\bm{v}}\vert$ such that
\begin{equation}\label{EQN:OrderUNITY}
\frac{|\vec{O}(\bm{p'}+\bm{v})-\vec{O}(\bm{p'})|}{|\vec{O}(\bm{p'})|}\sim \mathcal{O}(1).
\end{equation}

Note that not all changes in the (vector of) observables may be significant. In an example where one has a total of one observable, as explored in Sec.~\ref{SEC:PBHs}, an order-unity increase in the fraction of dark matter in the form of PBHs should not count as significant when assessing the scenario wherein a significant fraction of dark matter is in the form of PBHs (of course, a decrease would be significant). Or, as explored in Sec.~\ref{SEC:HabitableHalos}, an order-unity increase in the fraction of protons that end up in habitable dark-matter halos should not count as significant when what we are really interested in is the potential for such halos to give rise to life. Thus one needs to heed any such interpretation of the observables when applying Eq.~(\ref{EQN:OrderUNITY}).

If we denote by $\Delta[\bm{p'}; \hat{\bm{v}}]$, the size of the range of allowed parameter values, starting at $\bm{p'}$, in the direction $\hat{\bm{v}}$, then our global measure of fine-tuning, $\mathcal{G}(\bm{p'}; \hat{\bm{v}})$, is given by 
\begin{equation}\label{EQN:Global}
\mathcal{G}(\bm{p'}; \hat{\bm{v}})\equiv\log_{10}\left(\frac{\textcolor{black}{\Delta[\bm{p'}; \hat{\bm{v}}]}}{|\bm{v}|}\right).
\end{equation}
We introduce the convention that if an order-unity fractional change in the observable does not occur by the point at which one reaches the edge of parameter space, then $|\bm{v}|\equiv\Delta[\bm{p'}; \hat{\bm{v}}]$, in which case $\mathcal{G}(\bm{p'}; \hat{\bm{v}})=0$. Note then that this measure of global fine-tuning is non-negative, $\mathcal{G}(\bm{p'}; \hat{\bm{v}})\geqslant 0$.

Alternatively, one may be interested in the entire range of parameter values along the line through $\bm{p'}$, parallel to $\hat{\bm{v}}$. Then, denoting by $\Delta\left[\bm{p'}; \hat{\bm{v}}^{\pm}\right]$, the size of the range of allowed parameter values, starting at $\bm{p'}$, in either direction $\hat{\bm{v}}$ or  $-\hat{\bm{v}}$, we have
\begin{equation}\label{EQN:Global2Delta}
\Delta\left[\bm{p'}; \hat{\bm{v}}^{\pm}\right]\equiv \Delta\left[\bm{p'}; \hat{\bm{v}}\right]+\Delta\left[\bm{p'}; -\hat{\bm{v}}\right],
\end{equation}
where $\Delta\left[\bm{p'}; -\hat{\bm{v}}\right]$ is the size of the range of allowed parameter values, starting at $\bm{p'}$, in the direction $-\hat{\bm{v}}$. Let
\begin{equation}\label{EQN:Global2v}
|\bm{v}^{\pm}|\equiv|\bm{v}|+|\bm{v}^{-}|,
\end{equation}
where ${|\bm{v}^{-}|}$ is the size of the vector that leads to an order-unity fractional change of the observable in the direction $-\hat{\bm{v}}$; and where we invoke the convention introduced immediately after Eq.~(\ref{EQN:Global}), appropriately modified for the $-\hat{\bm{v}}$ direction. Then we can define another (manifestly non-negative) measure of global fine-tuning, which we will denote by $\tilde{\mathcal{G}}(\bm{p'}; \hat{\bm{v}}^{\pm})$,  namely, 
\begin{equation}\label{EQN:Global2}
\tilde{\mathcal{G}}(\bm{p'}; \hat{\bm{v}}^{\pm})\equiv\log_{10}\left(\frac{\textcolor{black}{\Delta\left[\bm{p'}; \hat{\bm{v}}^{\pm}\right]}}{|\bm{v}^{\pm}|}\right).\end{equation}

As in the case for the general local measure of fine-tuning, we distinguish orders of global fine-tuning (treating our two global measures independently), based on the numerical value of ${\mathcal{G}}(\bm{p'}; \hat{\bm{v}})$ or $\tilde{\mathcal{G}}(\bm{p'}; \hat{\bm{v}}^{\pm})$. In particular, we distinguish the following cases:
\begin{subequations}\label{EQN:GlobalOrdDEF}
\begin{align}
\mathcal{G}(\bm{p'}; \hat{\bm{v}}) = 0 &\equiv\textrm{Order }0,\label{EQN:GlobalOrderNo}\\
0<\mathcal{G}(\bm{p'}; \hat{\bm{v}})\leqslant \frac{1}{2} &\equiv\textrm{Order }1,\\
\frac{1}{2}<\mathcal{G}(\bm{p'}; \hat{\bm{v}})\leqslant 1 &\equiv\textrm{Order }2,\\
&\vdots\nonumber\\
\frac{N}{2}<\mathcal{G}(\bm{p'}; \hat{\bm{v}})\leqslant \frac{N+1}{2} &\equiv\textrm{Order }N+1, \label{EQN:GlobalOrderNplus1}
\end{align}
\end{subequations}
where $N\in\{0,1,\dots\}$ can, in principle, be unbounded. A similar set of cases  for $\tilde{\mathcal{G}}(\bm{p'}; \hat{\bm{v}}^{\pm})$ can be obtained by replacing $\mathcal{G}(\bm{p'}; \hat{\bm{v}})$ with $\tilde{\mathcal{G}}(\bm{p'}; \hat{\bm{v}}^{\pm})$ in Eqs.~(\ref{EQN:GlobalOrdDEF}).

We now apply the formalism developed above to two pertinent astrophysical settings.

\section{Dark matter as primordial black holes}\label{SEC:PBHs}

To illustrate how our measures of fine-tuning operate, we first characterize levels of local and global fine-tuning as they arise in models where a significant fraction of dark matter in the universe is in the form of PBHs. 

First posited about 50 years ago, PBHs are hypothetical entities thought to arise from large overdensities on horizon scales in the early universe~\cite{zeldovich+novikov_67, hawking_71, carr+hawking_74, carr_75}. Such overdensities can be sourced by an early inflationary phase (see, e.g., Refs.~\cite{garcia-bellido+al_96, clesse+garcia-bellido_15, garcia-bellido+morales_17, hertzberg+yamada_18}). There has been a resurgence of interest in their existence due to the discovery of massive ($\sim 30 M_{\odot}$) black holes by the LIGO-Virgo Collaboration~\cite{abbott+al_16}; as large numbers of black holes of such large masses are difficult to manufacture through stellar collapse (at least for solar metallicity environments)~\cite{belczynski+al_10, belczynski+al_16}. Massive PBHs have also been posited as seeds for the generation of supermassive black holes thought to reside in the centers of galaxies~\cite{rubin+al_01, bean+magueijo_02, duechting_04, clesse+garcia-bellido_15, carr+silk_18}. The idea that PBHs could constitute dark matter has been around for about 40 years~\cite{chapline_75}. It gained further attention as a result of the LIGO-Virgo results~\cite{bird+al_16}, and comprises a distinct alternative to particle-based models of dark matter (see Ref.~\cite{carr+al_16} for a review). 

To describe fine-tuning in models that claim that PBHs are a significant fraction of dark matter, we need to identify, in accord with Eq.~(\ref{EQN:ObservableMapping}), parameters that characterize the formation of PBHs, as well as observables whose finely tuned nature we are investigating. 

For the sake of simplicity, we fix the parameters to be those that arise in a specific  {\it extended} mass function for PBHs --- one that is thought to be a good approximation to the mass function that would arise assuming certain mechanisms for their production (e.g., those invoking cosmic inflation).\footnote{In models that source PBHs from, say, single-scalar-field models of cosmic inflation (see, e.g., Refs.~\cite{garcia-bellido+morales_17, hertzberg+yamada_18}), one could fix the parameters to be those that appear in a suitable scalar potential $V(\phi)$, together with initial conditions for the scalar field.} Also, we will restrict attention to the case where there is a single observable: the fraction of dark matter in the form of PBHs.

The extended mass function we invoke takes a lognormal form (see, e.g., Refs.~\cite{dolgov+silk_93, green_16, blinnikov+al_16, kannike+al_17, kuhnel+freese_17, carr+al_17}). In particular, the differential mass function of the fraction of dark matter in the form of PBHs, denoted by $f(M)$ (where $M$ is the present-day PBH mass), will be assumed to be
\begin{align}\label{EQN:LogNormal}
f(M)&\equiv\frac{1}{\Omega_{\textrm{DM}}}\frac{d\Omega_{\textrm{PBH}}}{d\ln M}\nonumber\\ &\equiv\frac{f_{\textrm{PBH}}}{\sqrt{2\pi}\sigma}\exp\left[-\frac{1}{2\sigma^2}\left(\ln M - \ln M_{c}\right)^2\right], 
\end{align}
where: $\Omega_{\textrm{DM}}\equiv\rho_{\textrm{DM}}/\rho_{\textrm{c}}$ is the ratio of the density of dark matter today to the critical density today; $\Omega_{\textrm{PBH}}\equiv\rho_{\textrm{PBH}}/\rho_{\textrm{c}}$ is the ratio of the density of matter in the form of PBHs today to the critical density today; $M_{c}$ is the `central' value of the mass of PBHs at which the lognormal distribution peaks; and $\sigma$ is the standard deviation of the distribution. Note also that 
\begin{equation}\label{EQN:fPBH}
f_{\textrm{PBH}} = \int_{-\infty}^{\infty} d\ln M f(M) = \frac{\Omega_{\textrm{PBH}}}{\Omega_{\textrm{DM}}},
\end{equation}
is the total fraction of dark matter in the form of PBHs. We thus have a two-parameter mass function, where the parameters are $(M_{c}, \sigma)$. In what follows, we will consider a dimensionless, scaled version of these parameters, namely $(\log_{10}(M_{c}/M_{\odot}), \sigma)$. 

\begin{figure*}
\begin{minipage}{.49\linewidth}
\subfloat[]{\includegraphics[scale=0.6]{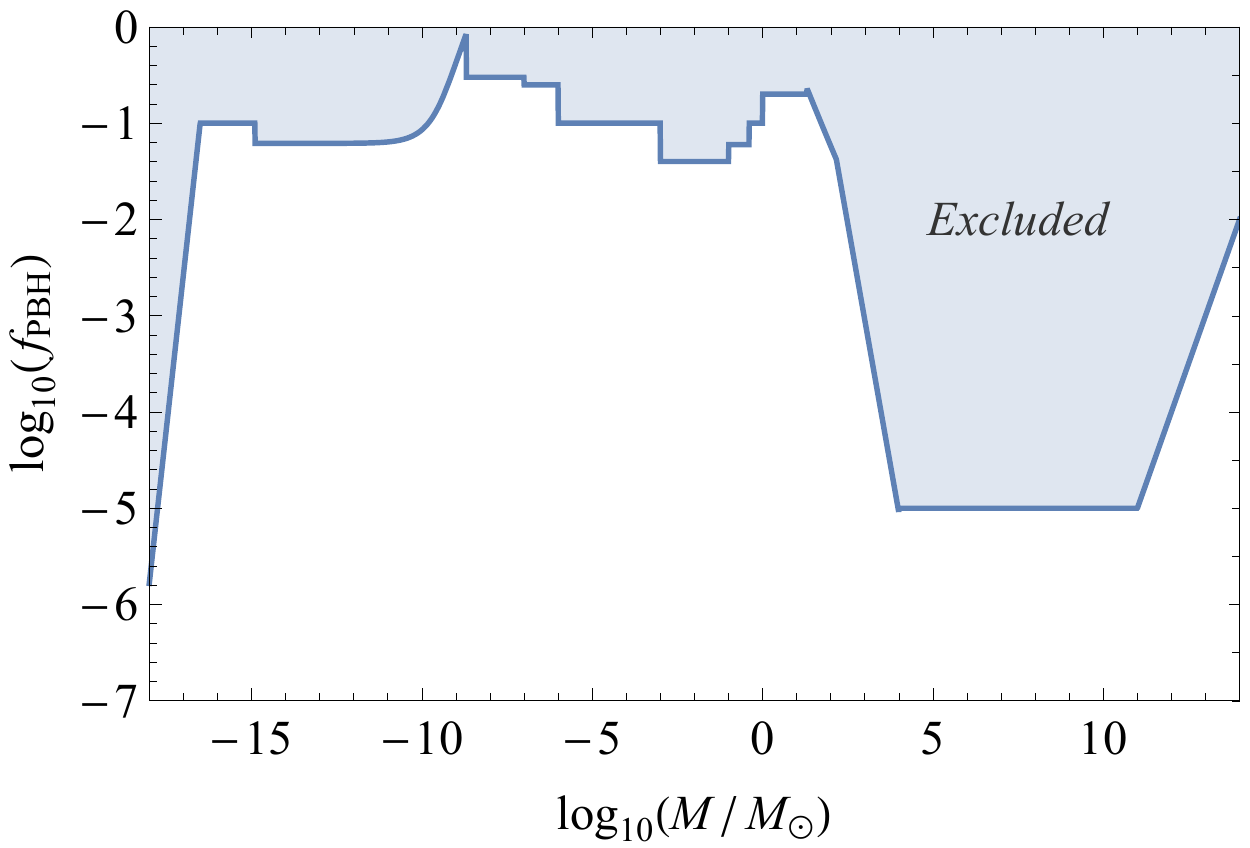}}
\end{minipage}
\begin{minipage}{.49\linewidth}
\subfloat[]{\includegraphics[scale=0.7]{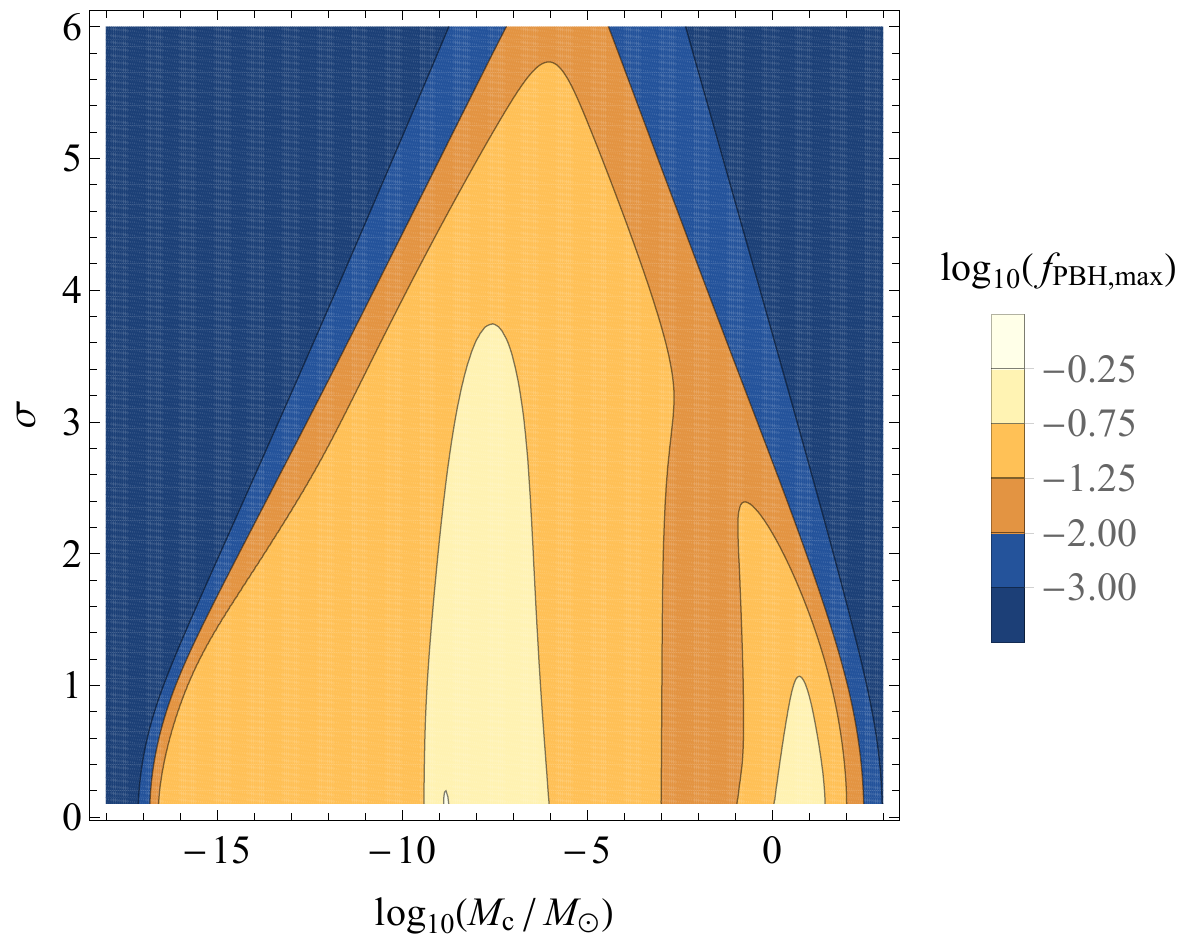}}
\end{minipage}
    \caption{(a) Monochromatic constraints on the fraction of dark matter in the form of PBHs, $f_{\textrm{PBH}}$, as a function of the mass, $M$, of PBHs today. A stylized subset of constraints from Refs.~\cite{josan+al_09, carr+al_16} has been employed to illustrate our fine-tuning formalism. The light-blue region denotes values of $f_{\textrm{PBH}}$ that are ruled out by observations and the dark-blue line demarcates the border of that region. (b) Maximal value of $f_{\textrm{PBH}}$ for the lognormal extended mass function in Eq.~(\ref{EQN:LogNormal}), such that the monochromatic constraints are satisfied [see Eq.~(\ref{EQN:fPBHmax})].}
    \label{FIG:PBHconstraints}
\end{figure*}

The observable we are interested in is related to $f_{\textrm{PBH}}$ in Eq.~(\ref{EQN:fPBH}) --- in particular, we will assume that the observable is the maximum allowed value of $f_{\textrm{PBH}}$, as determined by astrophysical constraints. As outlined in~\citet{carr+al_16}, astrophysical constraints on the abundance of PBHs have predominantly been derived assuming that the mass function of PBHs is monochromatic (i.e., the mass function is proportional to a Dirac delta function). The constraints come from a variety of considerations including: PBH evaporation, gravitational lensing experiments, various dynamical effects, and PBH accretion and radiation. These constraints are currently in flux (see Refs.~\cite{josan+al_09, carr+al_10, carr+al_16, carr+al_17, inomata+al_18} for a discussion of such constraints and for snapshots of how such constraints are employed in the literature). For the sake of our analysis, and in particular to determine the observable of interest for our fine-tuning analysis, we present an illustrative subset of such constraints in Fig.~\ref{FIG:PBHconstraints}(a). 

Our assumed differential mass function is extended, and we confront such a mass function with the monochromatic constraints displayed in Fig.~\ref{FIG:PBHconstraints}(a). Procedures have recently been developed that allow one to do this in a rigorous way (see Refs.~\cite{carr+al_17, inomata+al_17, inomata+al_18}). In particular, for each observable, the constraint that the extended mass function needs to satisfy is 
\begin{equation}\label{EQN:EMFconstraint}
\int d\ln M \frac{f(M)}{f_{\textrm{max}}^{\textrm{mono}}(M)}\leq 1, 
\end{equation}
where $f_{\textrm{max}}^{\textrm{mono}}(M)$ is the maximum value of the fraction of dark matter in the form of PBHs, under the assumption of a monochromatic mass function. We implement a simplified version of the full procedure, in which we assume that ${f_{\textrm{max}}^{\textrm{mono}}(M)}$ is the solid blue line in Fig.~\ref{FIG:PBHconstraints}(a) for the entire PBH mass-range of interest. Thus, substituting Eq.~(\ref{EQN:LogNormal}) into Eq.~(\ref{EQN:EMFconstraint}), and rearranging, we obtain an expression for the maximum value of the fraction of dark matter in the form of PBHs:
\begin{widetext}
\begin{equation}\label{EQN:fPBHmax}
f_{\textrm{PBH,max}}=\left\{\int d\ln M\frac{1}{f_{\textrm{max}}^{\textrm{mono}}(M)}\frac{1}{\sqrt{2\pi}\sigma}\exp\left[-\frac{1}{2\sigma^2}\left(\ln M - \ln M_{c}\right)^2\right]\right\}^{-1}.
\end{equation}
\end{widetext}

We now expressly connect this scheme for analyzing PBHs with the general setting in which we derived measures for fine-tuning (in Sec.~\ref{SEC:Measures}). In particular, we choose: $\bm{p}\to (\log_{10}(M_{c}/M_{\odot}), \sigma)$ and ${O}\to f_{\textrm{PBH,max}}$. 

Performing the integral in Eq.~(\ref{EQN:fPBHmax}) numerically, we find the result displayed in Fig.~\ref{FIG:PBHconstraints}(b). The shape of the region where one obtains a significant maximal fraction, e.g., $\log_{10}(f_{\textrm{PBH,max}}) \gtrsim -3$ is roughly symmetric due to the symmetry of the lognormal mass function (and the approximate symmetry of the monochromatic constraints). In addition, for a fixed $M_{c} $, we find that increasing the width of the mass function, viz., increasing $\sigma$, leads to an eventual decrease in $\log_{10}(f_{\textrm{PBH,max}})$, indicating that generically, one cannot avoid the constraints by flattening out the mass function. Finally, we note a small island around $\log_{10}(M_{c}/M_{\odot})\sim -8.8$ and $\sigma\sim 0.2$, where one obtains a significant fraction of dark matter in the form of PBHs, viz., $f_{\textrm{PBH,max}} \gtrsim 50\%$. This island corresponds to the permissive nature of the selected constraints around the same PBH mass in Fig.~\ref{FIG:PBHconstraints}(a). 

\begin{figure*}
\begin{minipage}{.49\linewidth}
\centering
\subfloat[]{\includegraphics[scale=0.65]{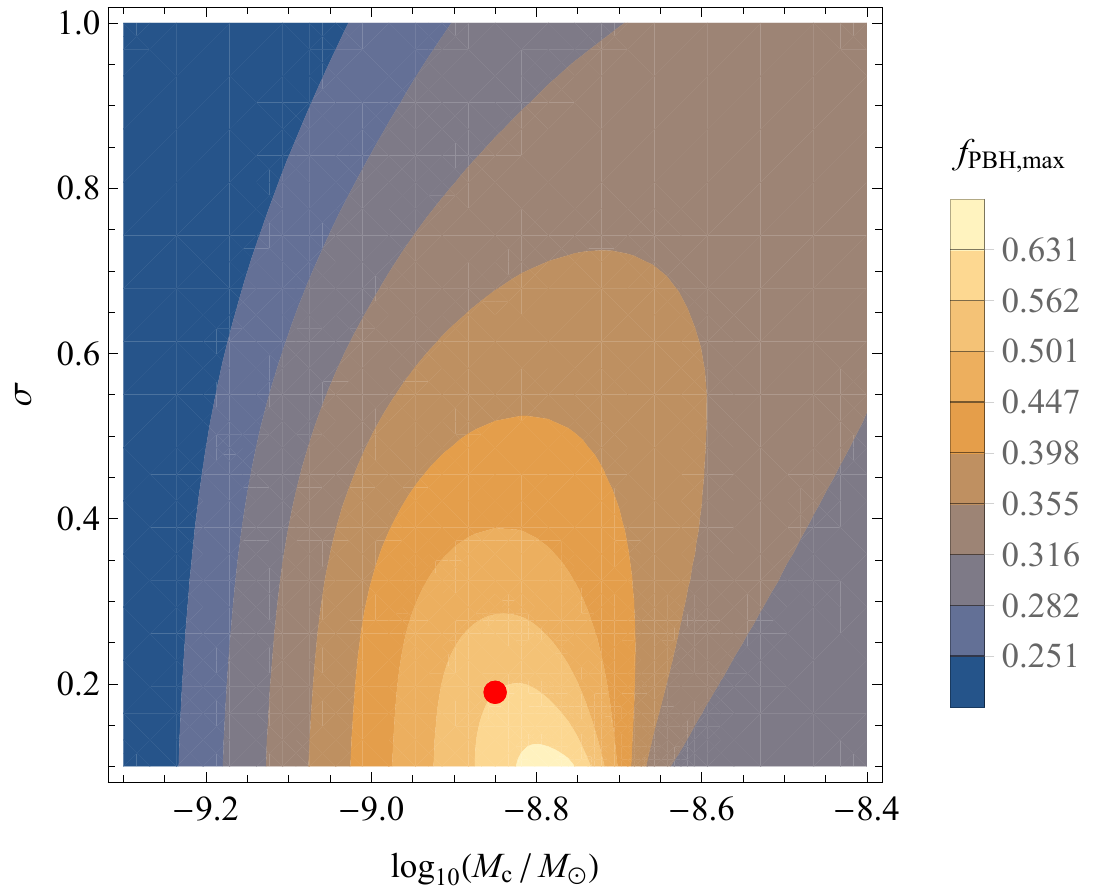}}
\end{minipage}
\begin{minipage}{.49\linewidth}
\centering
\subfloat[]{\includegraphics[scale=0.65]{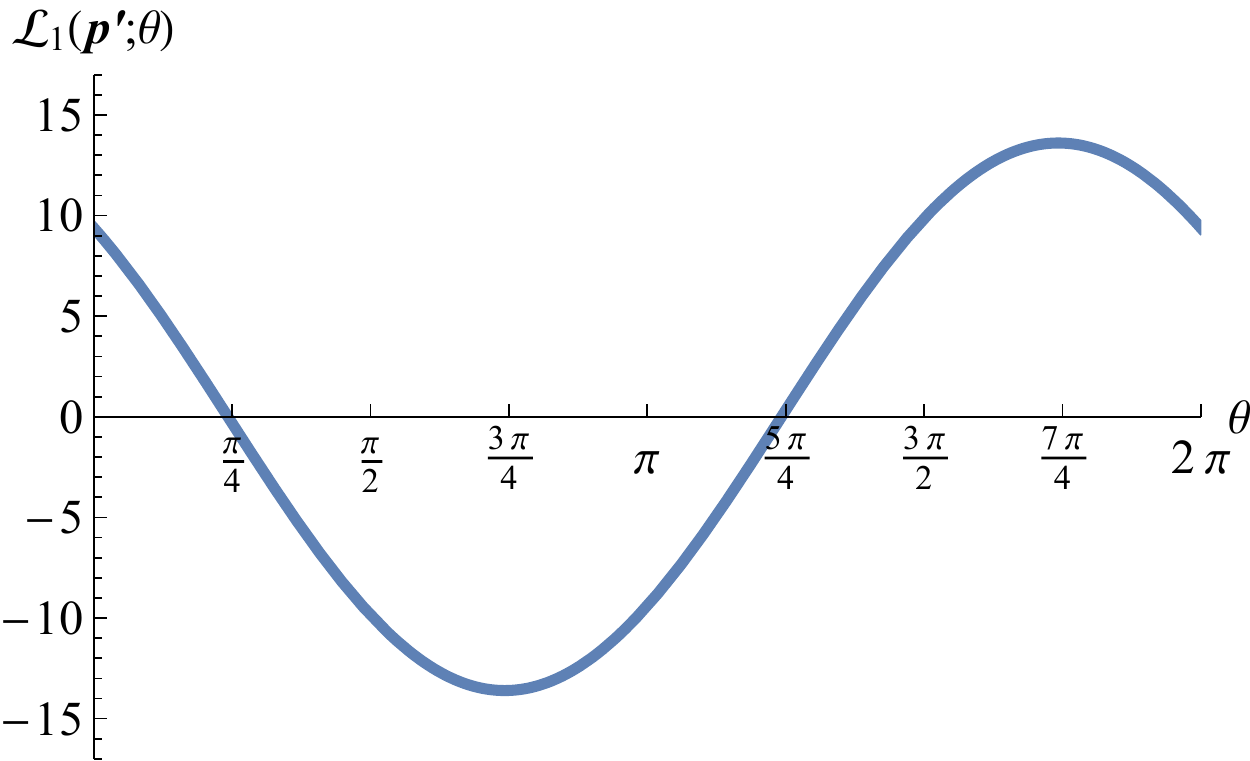}}
\end{minipage}
\begin{minipage}{.49\linewidth}
\centering
\subfloat[]{\includegraphics[scale=0.65]{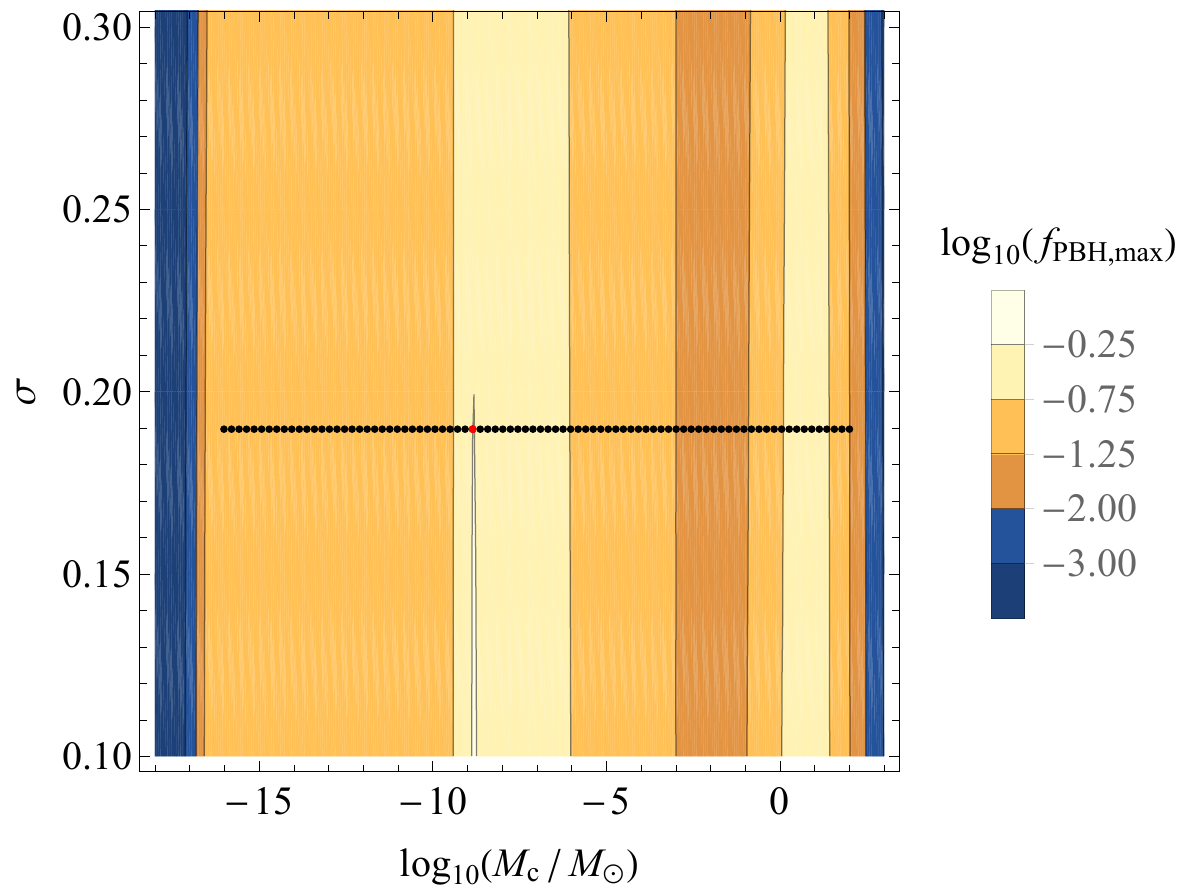}}
\end{minipage}
\begin{minipage}{.49\linewidth}
\centering
\subfloat[]{\includegraphics[scale=0.65]{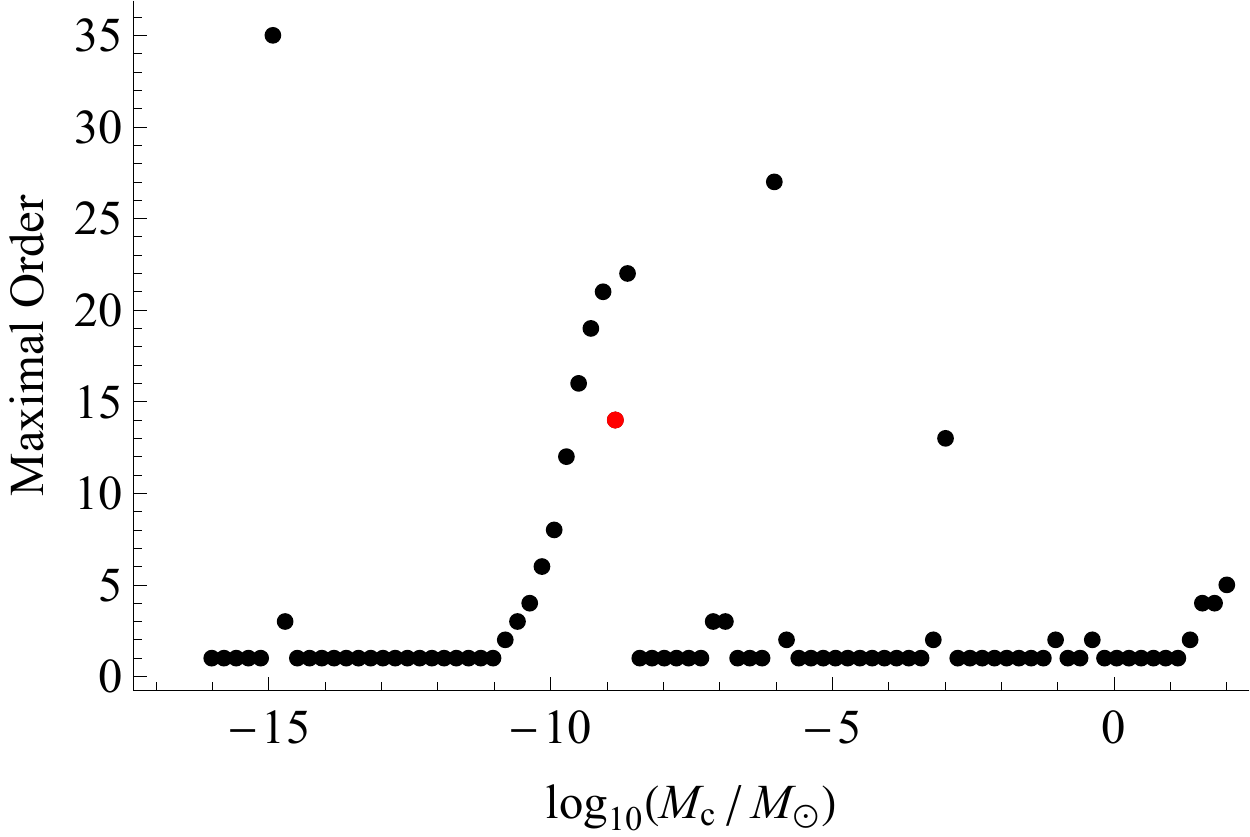}}
\end{minipage}
    \caption{The measure of local fine-tuning for the scenario wherein PBHs make-up a significant fraction of dark matter. (a) A close-up of the corresponding region from Fig.~\ref{FIG:PBHconstraints}(b), showing the maximal value of $f_{\textrm{PBH}}$ for the lognormal extended mass function in Eq.~(\ref{EQN:LogNormal}), such that the monochromatic constraints are satisfied [see Eq.~(\ref{EQN:fPBHmax})]. The red dot corresponds to a particular point in parameter space, $\bm{p'}=(-8.85,0.190)$, at which measures of local (and global) fine-tuning have been calculated. (b) The level of local fine-tuning at point $\bm{p'}=(-8.85,0.190)$, as a function of direction, $\theta$, in parameter space. As usual, $\theta$ is measured anticlockwise from the positive $x$-axis in parameter space. We have plotted $\mathcal{L}_{1}(\bm{p'};\theta)$ as it appears in Eq.~(\ref{EQN:Lform1}). (c) Selected points (black and red) in parameter space at which the maximal order of local fine-tuning was computed. The red dot corresponds to $\bm{p'}=(-8.85,0.190)$ as in (a). (d) The maximal order of local fine-tuning at each corresponding point in (c). The red dot in (d) yields the maximal order of local fine-tuning at $\bm{p'}=(-8.85,0.190)$ --- viz., Order 14.
        }\label{FIG:LocalFTPBH}
\end{figure*}

We are now in a position to compute the  level of local fine-tuning in such a model using Eq.~(\ref{EQN:Lform1}). In particular, we are interested in the scenario in which a significant fraction of dark matter is in the form of PBHs, and so we assume that the point in parameter space we are primarily interested in [viz., ${\bm p'}$ in Eq.~(\ref{EQN:Lform1})] corresponds to a large $f_{\textrm{PBH,max}}$. We choose $\log_{10}(f_{\textrm{PBH,max}}) \sim -0.25$, and thereby characterize fine-tuning in a scenario where about $56\%$ of the total density of dark matter is in the form of PBHs. Our chosen point in parameter space with this property is marked by the red dot in Fig.~\ref{FIG:LocalFTPBH}(a) [it has coordinates $\bm{p'}=(-8.85,0.190)$].  

In applying Eq.~(\ref{EQN:Lform1}) we furthermore note that our parameter space is two-dimensional, so we can write $\hat{\bm{\epsilon}}=(\cos\theta, \sin\theta)$. Thus, once we have fixed $\bm{p'}$, the level of local fine-tuning, $\mathcal{L}(\bm{p'};\theta)$ [or, alternatively, $\mathcal{L}_{1}(\bm{p'};\theta)$], is purely a function of the angle $\theta$. We plot $\mathcal{L}_{1}(\bm{p'};\theta)$ in Fig.~\ref{FIG:LocalFTPBH}(b), and find a maximal level of local fine-tuning of Order 14. According to the classification scheme defined in Eqs.~(\ref{EQN:LocalClassif}), this corresponds to a significant level of local fine-tuning. We note also, as anticipated in the final paragraph of Sec.~\ref{SEC:LocalMeasure}, there are directions in parameter space where there is no local fine-tuning [where $\mathcal{L}(\bm{p'};\theta)=0=\mathcal{L}_{1}(\bm{p'};\theta)$] --- such directions correspond to those in which $\hat{\bm{\epsilon}}$ points along a contour of the observable $f_{\textrm{PBH,max}}$. 

To provide some context for the levels of fine-tuning quoted for the chosen point $\bm{p'}$, we compute the maximal order of local fine-tuning at a set of points with the same mass-function width $\sigma$, but with different values of the mass, $M_{c}$, at which the mass function peaks. The chosen points are displayed in Fig.~\ref{FIG:LocalFTPBH}(c). The maximal order of local fine-tuning for these points is displayed in Fig.~\ref{FIG:LocalFTPBH}(d). Most selected points display maximal orders that are low compared to the chosen point $\bm{p'}$.

To compute the level of global fine-tuning (relative to the same point in parameter space, ${\bm p'}$) in accord with the scheme introduced in Sec.~\ref{SEC:GlobalMeasure}, we need to fix some finite range over which parameters can take values. In what follows, we fix this range to correspond to: $M_{c} \in [5\times 10^{14}\textrm{g}, 10^{4}M_{\odot}]$ and $\sigma \in [0.1,10]$. The lower bound on $M_{c}$  approximates the mass of black holes at formation that would have evaporated, via Hawking radiation, by the present epoch~\cite{carr+al_10, carr+al_17}, whereas the upper limit is a conservative, illustrative upper bound based on astrophysical constraints [see Fig.~\ref{FIG:PBHconstraints}(a)].\footnote{See Ref.~\cite{blinnikov+al_16} for a discussion of one mechanism that allows for such an upper cutoff on the mass of PBHs today.} The lower bound on the range of $\sigma$ is chosen such that one does indeed obtain an extended mass function, and the upper bound is chosen so that we subsequently probe at least two orders of magnitude in $\sigma$. 

\begin{figure*}
\begin{minipage}{.325\linewidth}
\centering
\subfloat[]{\includegraphics[scale=0.5]{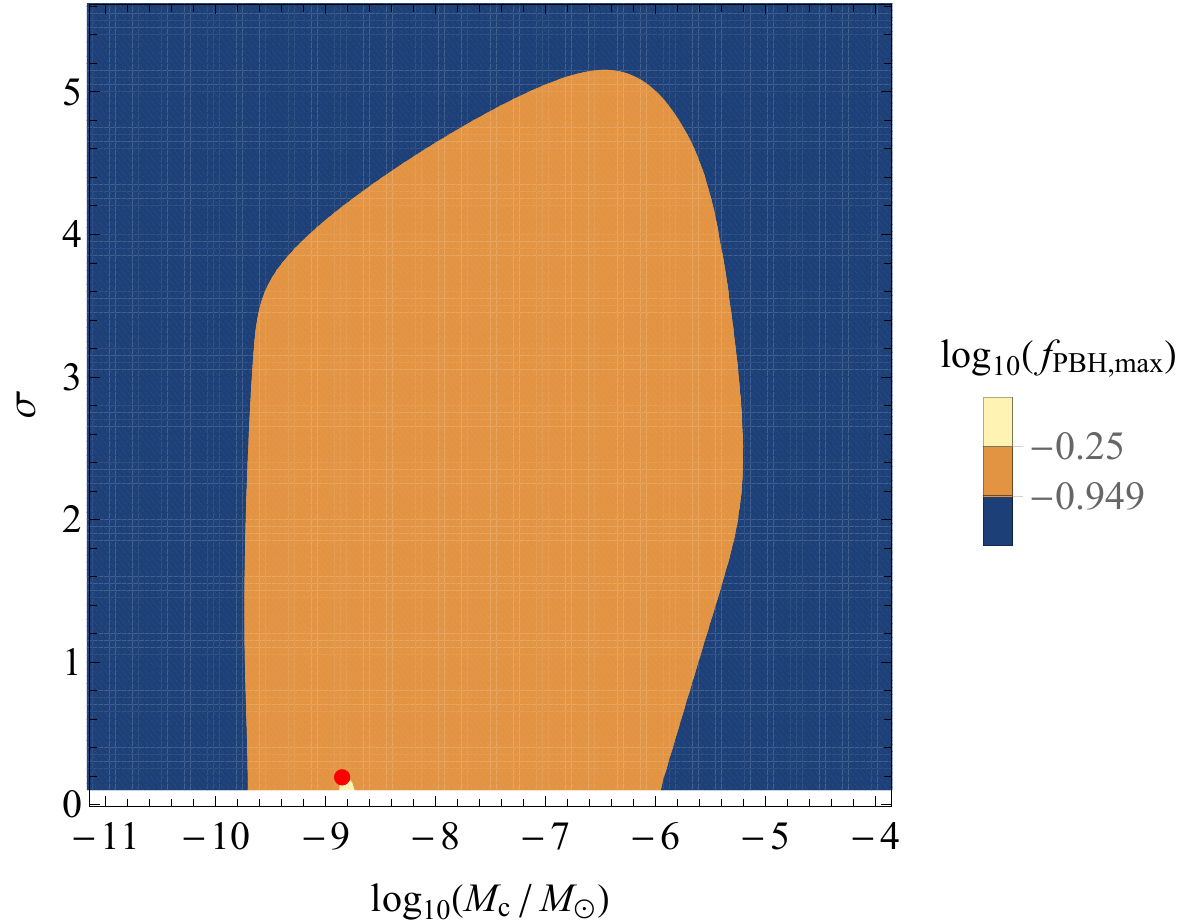}}
\end{minipage}
\begin{minipage}{.325\linewidth}
\centering
\subfloat[]{\includegraphics[scale=0.44]{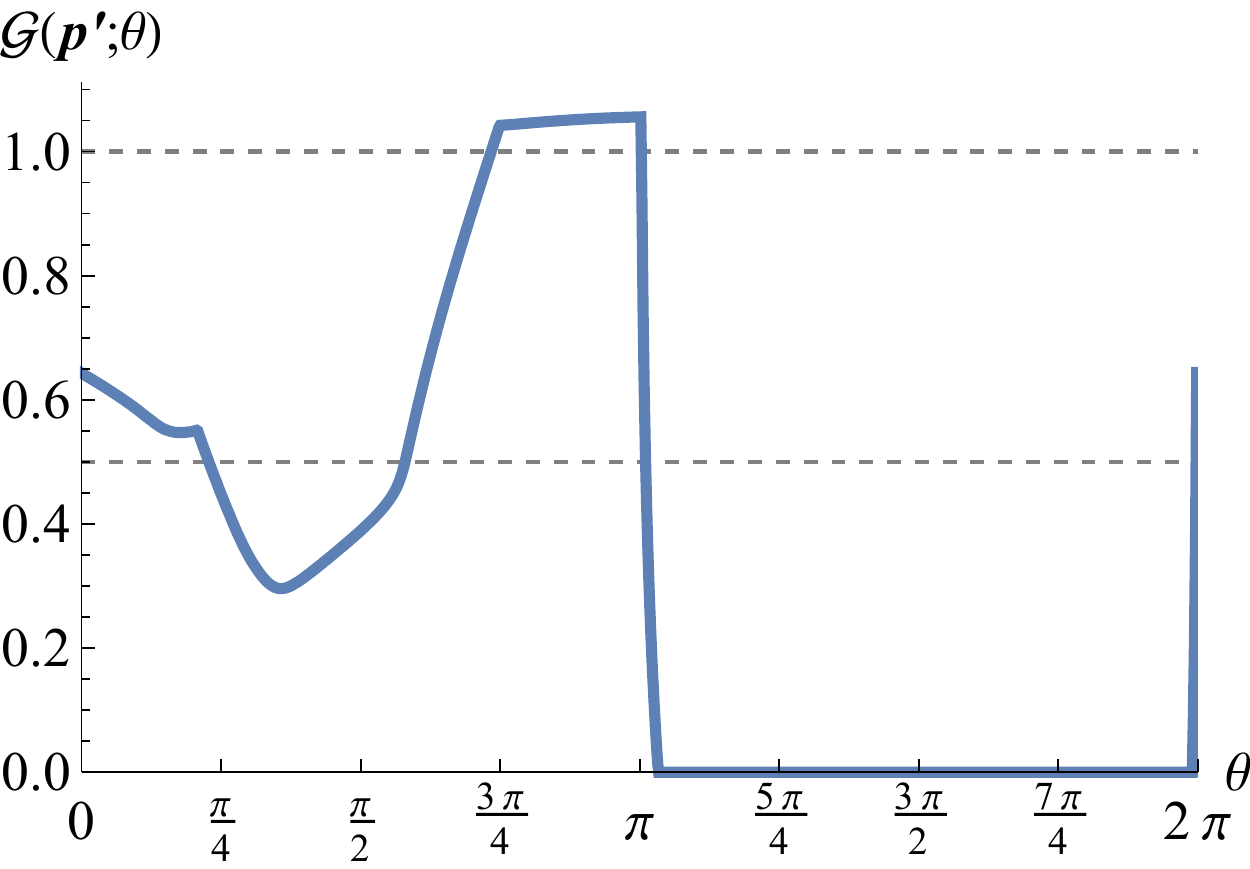}}
\end{minipage}
\begin{minipage}{.325\linewidth}
\centering
\subfloat[]{\includegraphics[scale=0.44]{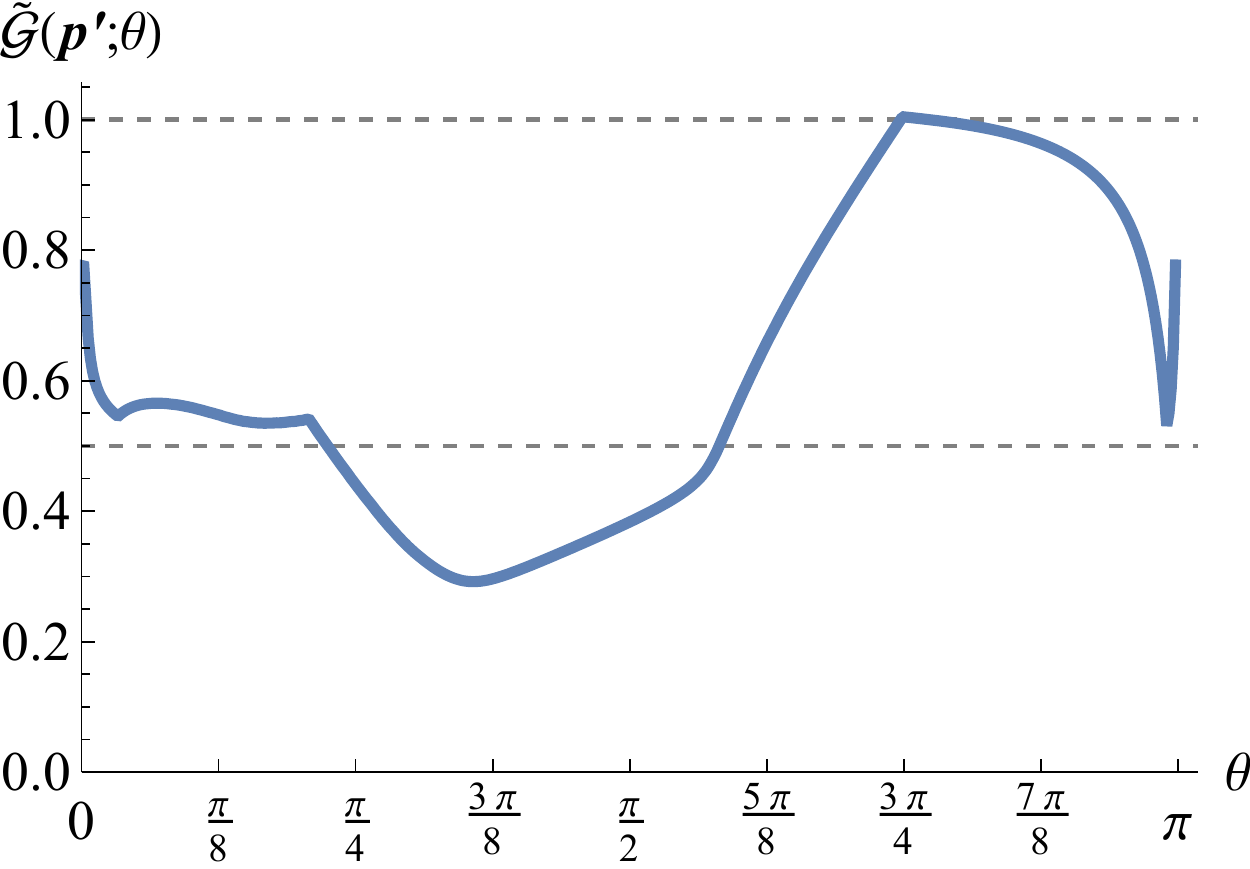}}
\end{minipage}
    \caption{Measures of global fine-tuning for the scenario wherein PBHs make up  a significant fraction of dark matter. (a) A close-up of the corresponding region from Fig.~\ref{FIG:PBHconstraints}(b), showing the maximal value of $f_{\textrm{PBH}}$ consistent with the observational constraints displayed in Fig.~\ref{FIG:PBHconstraints}(a). The red dot corresponds to a particular point in parameter space, $\bm{p'}=(-8.85,0.190)$, at which measures of global (and local) fine-tuning have been calculated. The dark-blue region corresponds to those points in parameter space that yield a significant change in $f_{\textrm{PBH,max}}$ as compared with the value of $f_{\textrm{PBH,max}}$ evaluated at $\bm{p'}$. (b) A computation of the measure of global fine-tuning in Eq.~(\ref{EQN:Global}). The dashed gray lines demarcate boundaries where the order of global fine-tuning changes [see Eqs.~(\ref{EQN:GlobalOrdDEF})]. As expected, $\mathcal{G}({\bm p';\theta})=0$ in directions that reach the edge of parameter space before a significant order-unity change in $f_{\textrm{PBH,max}}$ occurs. (c) A computation of the measure of global fine-tuning in Eq.~(\ref{EQN:Global2}).}\label{FIG:GlobalFTPBH}
\end{figure*}

Measures of global fine-tuning can then be computed using Eqs.~(\ref{EQN:OrderUNITY})--(\ref{EQN:Global2}). Again, due to the two-dimensional nature of the parameter space, directions in parameter space (labeled by, for example, $\hat{\bm v}$ in Sec.~\ref{SEC:GlobalMeasure}) can be parameterized by $\theta$. For each direction $\theta$ we may now use Eq.~(\ref{EQN:OrderUNITY}) to find the length of the vector that yields an order-unity change in the observable, $f_{\textrm{PBH,max}}$. Note, as discussed in Sec.~\ref{SEC:GlobalMeasure}, for the scenario that we are assessing, wherein a significant fraction of dark matter is in the form of PBHs, an order-unity increase in $f_{\textrm{PBH,max}}$ does not count as significant, and so we only probe order-unity decreases in $f_{\textrm{PBH,max}}$. We choose the order-unity change to correspond to $0.8$, and display numerical results for our computation of $\mathcal{G}({\bm p'};\theta)$ and $\tilde{\mathcal{G}}({\bm p'};\theta)$ in Fig.~\ref{FIG:GlobalFTPBH}.

In Fig.~\ref{FIG:GlobalFTPBH}(b), we see that the maximal level of global fine-tuning is of Order 3 [i.e., when $1< \mathcal{G}({\bm p';\theta}) \leq 3/2$ --- according to Eqs.~(\ref{EQN:GlobalOrdDEF})], and that $\mathcal{G}({\bm p';\theta})=0$ when the direction of interest does not yield a significant order-unity change in the observable before one reaches the edge of parameter space. When we take into account the entire range of parameter values along a line through ${\bm p'}$, as for the computation of $\tilde{\mathcal{G}}({\bm p';\theta})$, we find that the minimal level of global fine-tuning is Order 1. In sum: the hypothesis that PBHs make up a significant fraction of dark matter can indeed require high levels of both local and global fine-tuning.

\section{Habitable dark-matter halos from inflation}\label{SEC:HabitableHalos}

We turn now to our second illustrative example, that treats a less speculative scenario, wherein we compute levels of local and global fine-tuning in astrophysical models that estimate the fraction of protons in habitable dark-matter halos. Note that throughout this section, we will use `extended Planck units' (as in~\citet{tegmark+al_06}), where: $c=\hbar=G=k_{b}=|q_{e}|=1$, so that the reduced Planck mass is given by $\Mpl=(8\pi)^{-1/2}$.

We implement a simplified version of the scheme introduced in Ref.~\cite{tegmark+al_06}, which relates the fraction of protons, $f_{h}$ (`$h$' for `halo'), that end up in dark-matter halos whose densities lie in some specified range, to cosmological parameters: in particular, to the density of dark energy, $\rho_{\Lambda}$, and the amplitude of primordial density perturbations, $Q$ --- so that $f_{h} \equiv f_{h}(\rho_{\Lambda},Q)$.\footnote{See, for further details,~\citet[Sec. IVA]{tegmark+al_06}. Their final expression for $f_{h}$ [their Eq.~(50)] is also, in effect, a function of $\xi$, the matter density per photon-number density --- but we set this quantity to its present-day value quoted in Ref.~\cite{tegmark+al_06} for the sake of simplicity (viz., $\xi = 3.3 \times 10^{-28}$). Any dependence on $\xi_{\textrm{b}}$, the baryonic matter density per photon-number density, will be suppressed due to hard cutoffs employed for the density range over which such halos form.}  In particular, the standard Press-Schechter formalism is employed to compute the fraction of matter that has collapsed into dark-matter halos (of at least a certain mass at a certain time)~\cite{press+schechter_74}. Habitable halos correspond to those that lie within some mass-density range, viz., $\rho_{\textrm{min}} \leq \rho_{h}\leq\rho_{\textrm{max}}$ (where we set, as in~\cite{tegmark+al_06}, $\rho_{\textrm{min}}\equiv 10^{-128}$ and $\rho_{\textrm{max}}\equiv 10^{-120}$). They are `habitable' in the sense that the halos satisfy necessary conditions for the existence of life. From now on, $f_h$ will expressly refer to the fraction of protons in such habitable halos. 

Suitable values for the cosmological parameters, $(\rho_{\Lambda},Q)$, can be derived from a variety of models of cosmic inflation. We thus extend the discussion in Ref.~\citep{tegmark+al_06} to include an analysis where these parameters indeed arise from an early inflationary phase, sourced by a minimally coupled, single scalar field $\phi$, subject to a power-law potential, $V(\phi)$ (see, e.g., Ref.~\cite{hartle+hertog_13}). In particular, we choose
\begin{equation}\label{EQN:Vphi}
V(\phi)= \rho_{\Lambda}+\lambda\phi^{n},
\end{equation}
for specific values of $n$.

Assuming the slow-roll approximation, it is straightforward to derive $Q$ in terms of the potential (and its derivative with respect to $\phi$) when cosmological scales left the horizon during inflation: roughly $N_{*}=50$ to $60$ $e$-folds before the end of inflation (see, e.g., Ref.~\cite{planck_13INF}). One finds, to leading-order in slow-roll parameters, that
\begin{equation}\label{EQN:Qstar}
Q \sim \frac{1}{\sqrt{12\pi^{2}\Mpl^{4-n}}}\left(\frac{\lambda}{n^2}\right)^{1/2}\left(2 n N_{*} + \frac{n^2}{2}\right)^{(n+2)/{4}}.
\end{equation}
Here we have also assumed that $\rho_{\Lambda} \ll \lambda\phi^{n}$ for $\phi = \phi_{\textrm{end}}\sim n \Mpl/\sqrt{2}$, corresponding to the field value when slow-roll inflation ends, for the case where the field rolls down the potential starting from positive values of $\phi$. Thus, our observable of interest, the fraction of protons in habitable dark-matter halos, will now be expressed with the functional dependence: $f_{h}(\rho_{\Lambda},\lambda)$. 

\begin{figure}
\includegraphics[scale=0.9]{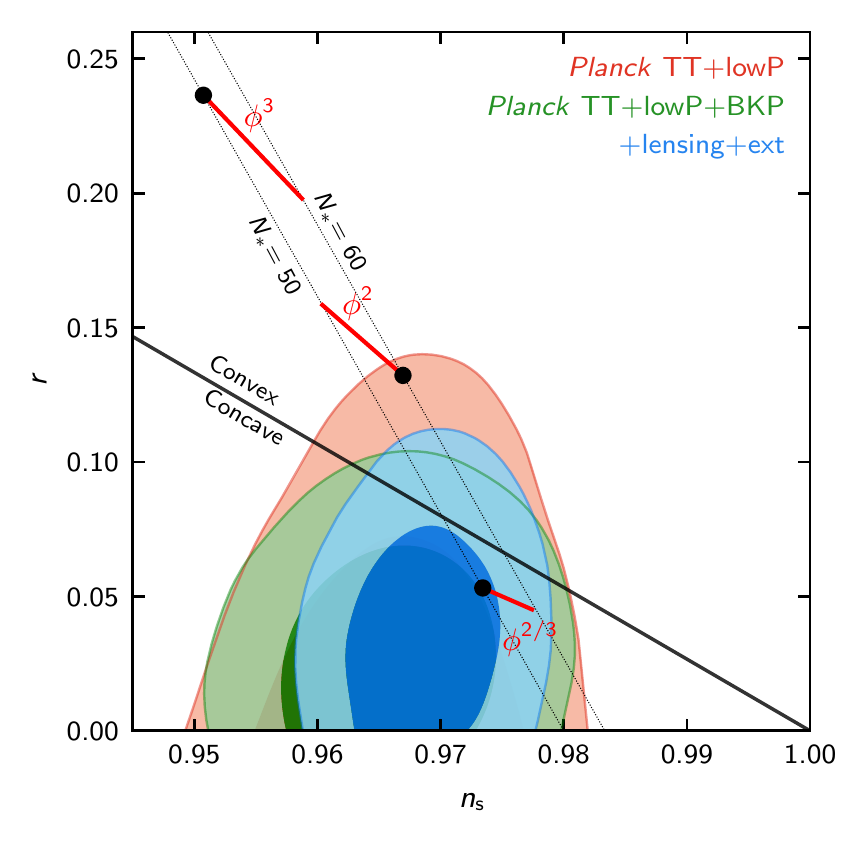}
\caption{ Relative degrees of agreement of the three inflationary potentials studied herein with results from the Planck Collaboration~\cite{ade+al_15}. The potentials, whose functional form is given in full in Eq.~(\ref{EQN:Vphi}), are labeled by their $\phi$-dependence. Black dots indicate the values of $N_{*}$ that we employ in our analysis.}
    \label{FIG:PlanckPlot}
\end{figure}

We focus on three values of the exponent, $n$, in Eq.~(\ref{EQN:Vphi}), namely, $n=2/3, 2$, and $3$. To leading-order in slow-roll parameters, and again assuming that $\rho_{\Lambda} \ll \lambda\phi_{\textrm{end}}^{n}$, one can compute the primordial scalar spectral index, $n_{s}$, and the tensor-to-scalar ratio, $r$, when cosmological scales left the horizon during inflation. One finds:
\begin{equation}
n_s   = \frac{4 N_{*}-n-4}{4 N_{*}+n},\qquad
r  = \frac{16n}{4 N_{*}+n}\label{EQN:nsrPL}.
\end{equation}
As shown in Fig.~\ref{FIG:PlanckPlot}, the three values of $n$ that we analyze lead to inflationary scenarios that have varying degrees of agreement with recent results from the Planck Collaboration~\cite{ade+al_15}, which has measured $n_s  = 0.968\pm 0.006$, and 
$r  < 0.11$. (Such qualitative differences in agreement between the three inflationary models we study do not change if we adopt the updated constraint $r<0.09$~\cite{bicep2+keck_16}.) For each value of $n$, the values of $n_s$ and $r$ plotted in Fig.~\ref{FIG:PlanckPlot} employ Eq.~(\ref{EQN:nsrPL}), where $N_{*}$ varies between 50 and 60 $e$-folds.

Now, for a fixed $n$, we connect the above scheme for analyzing the fraction of protons in habitable dark-matter halos with the general setting in which we derived measures for fine-tuning (in Sec.~\ref{SEC:Measures}). In particular, we choose $\bm{p}\to (\log_{10}\rho_{\Lambda}, \log_{10}\lambda)$ and ${O}\to f_{h}$. 

Moreover, in this example (in contrast to the example developed in Sec.~\ref{SEC:PBHs}), we are able to identify a point in the resulting two-dimensional parameter space that is consistent with results obtained by the Planck Collaboration~\cite{planck_15_CosmologicalParameters}. We find: $\rho_{\Lambda, {\rm obs}} \approx 1.16 \times 10^{-123}$ and $Q_{\rm obs}\equiv\sqrt{A_{s, {\rm obs}}} \approx 4.6\times 10^{-5}$ (where $A_{s, {\rm obs}}$ is the observed amplitude of primordial scalar perturbations). Then, for any $n$ and a fixed value of $N_{*}$, we can invert Eq.~(\ref{EQN:Qstar}) to obtain a value for $\lambda$, which we will denote by $\lambda_{\rm obs}(n , N_*)$, which is consistent with these results. We find, for the particular values of $N_*$ we will indeed use below: $\log_{10}[\lambda_{\rm obs}(2/3 , 50)]=-11.72$, $\log_{10}[\lambda_{\rm obs}(2 , 60)]=-12.16$, and $\log_{10}[\lambda_{\rm obs}(3 , 50)]=-12.55$. 

\begin{figure*}
\begin{minipage}{.49\linewidth}
\centering
\subfloat[]{\includegraphics[scale=0.45]{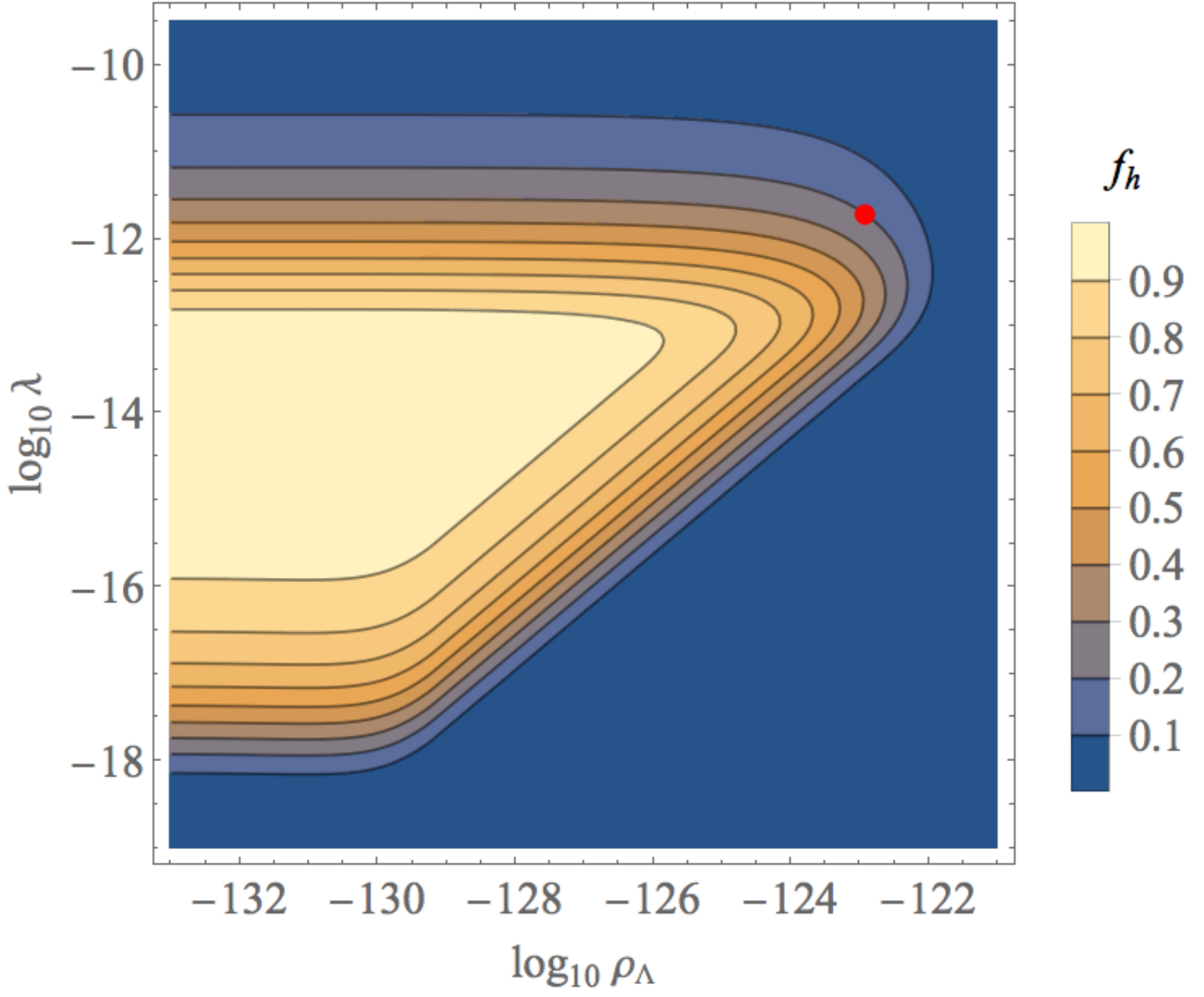}}
\end{minipage}
\begin{minipage}{.49\linewidth}
\centering
\subfloat[]{\includegraphics[scale=0.65]{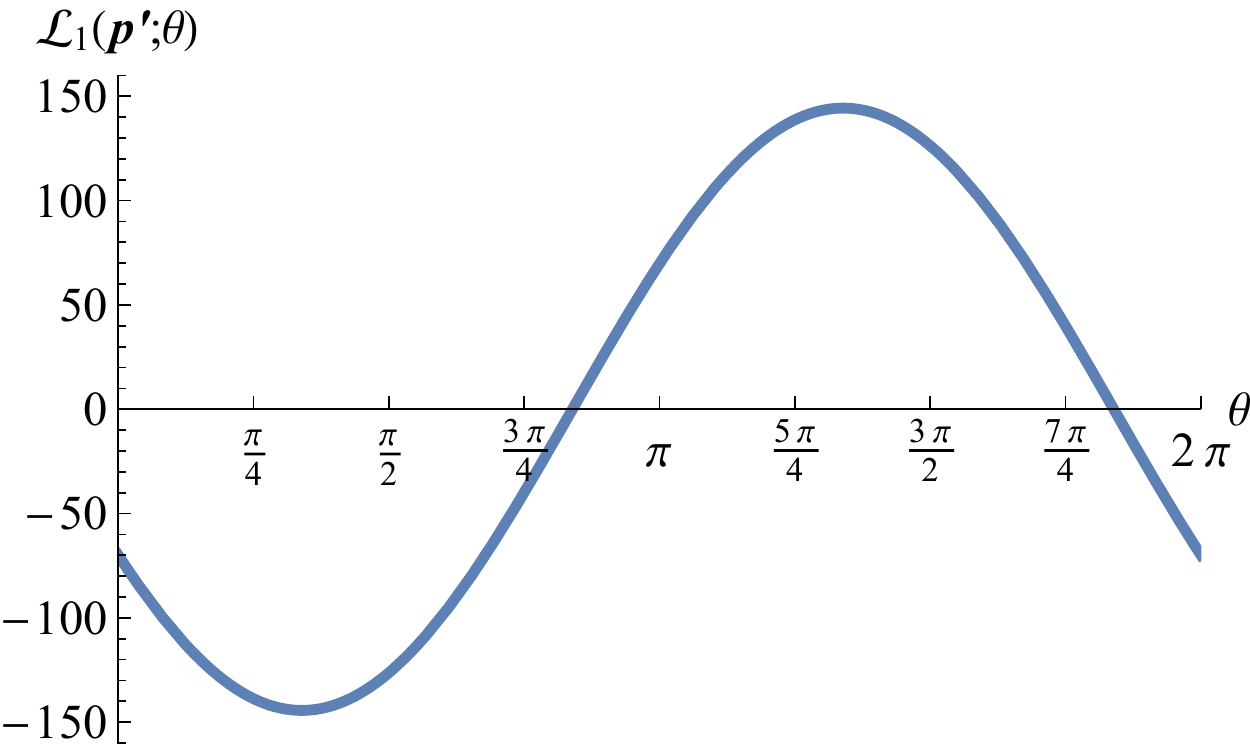}}
\end{minipage}
\begin{minipage}{.49\linewidth}
\centering
\subfloat[]{\includegraphics[scale=0.45]{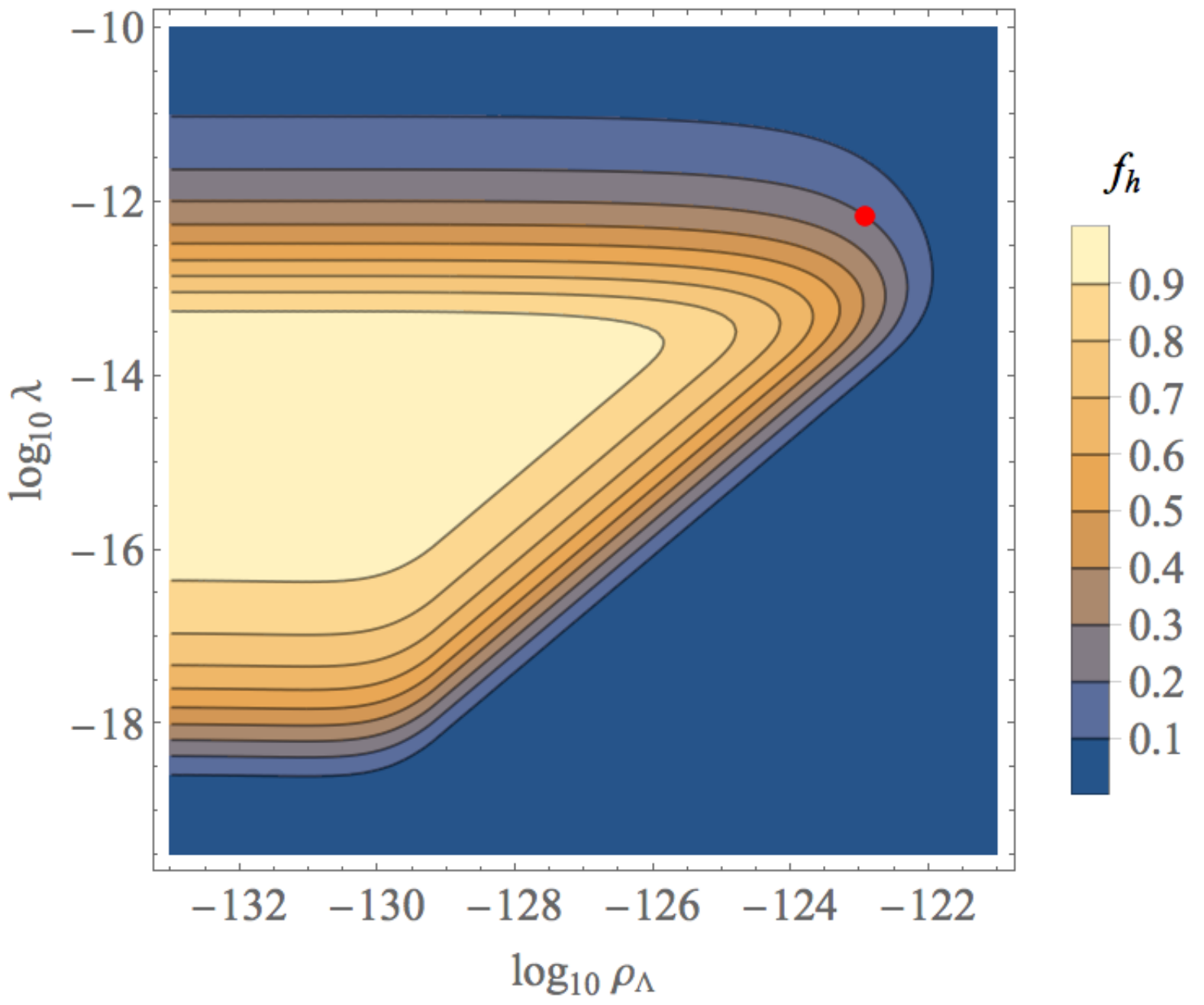}}
\end{minipage}
\begin{minipage}{.49\linewidth}
\centering
\subfloat[]{\includegraphics[scale=0.65]{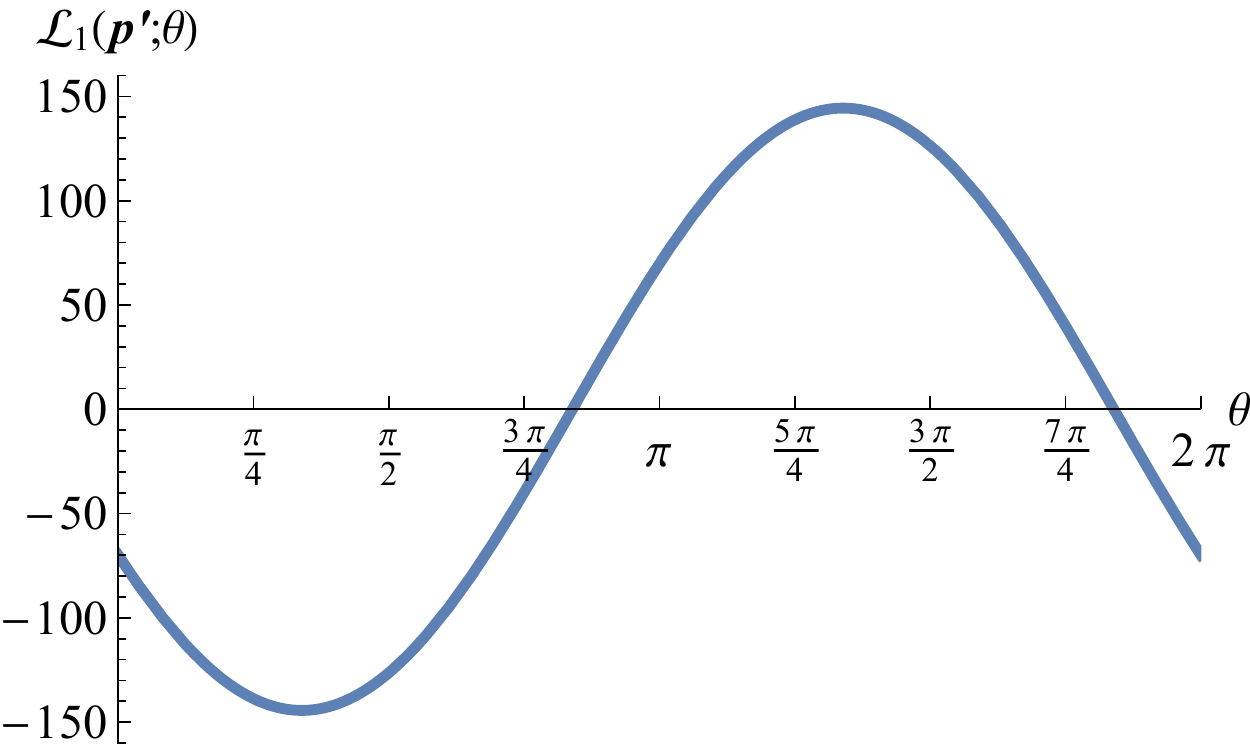}}
\end{minipage}
\begin{minipage}{.49\linewidth}
\centering
\subfloat[]{\includegraphics[scale=0.45]{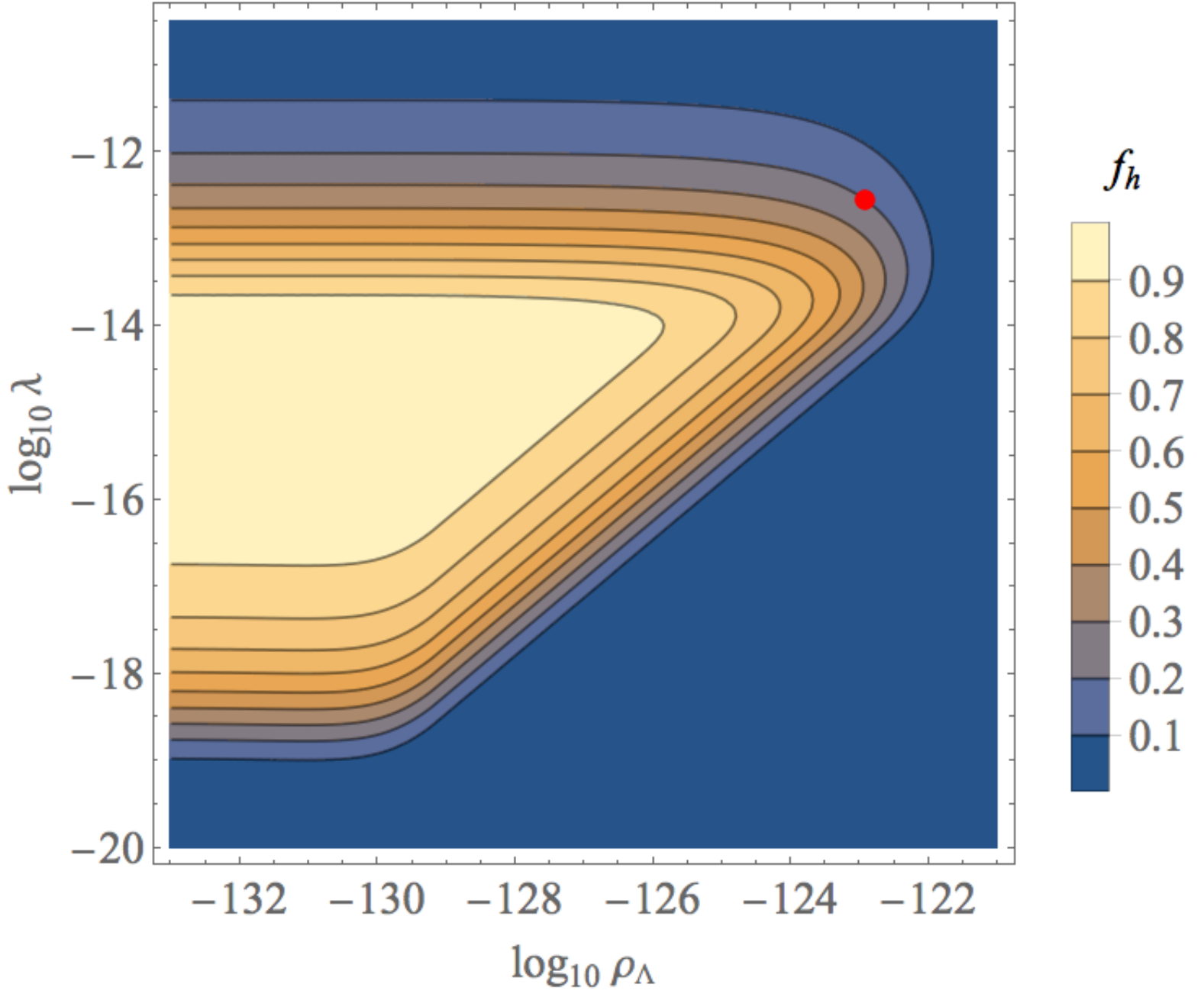}}
\end{minipage}
\begin{minipage}{.49\linewidth}
\centering
\subfloat[]{\includegraphics[scale=0.65]{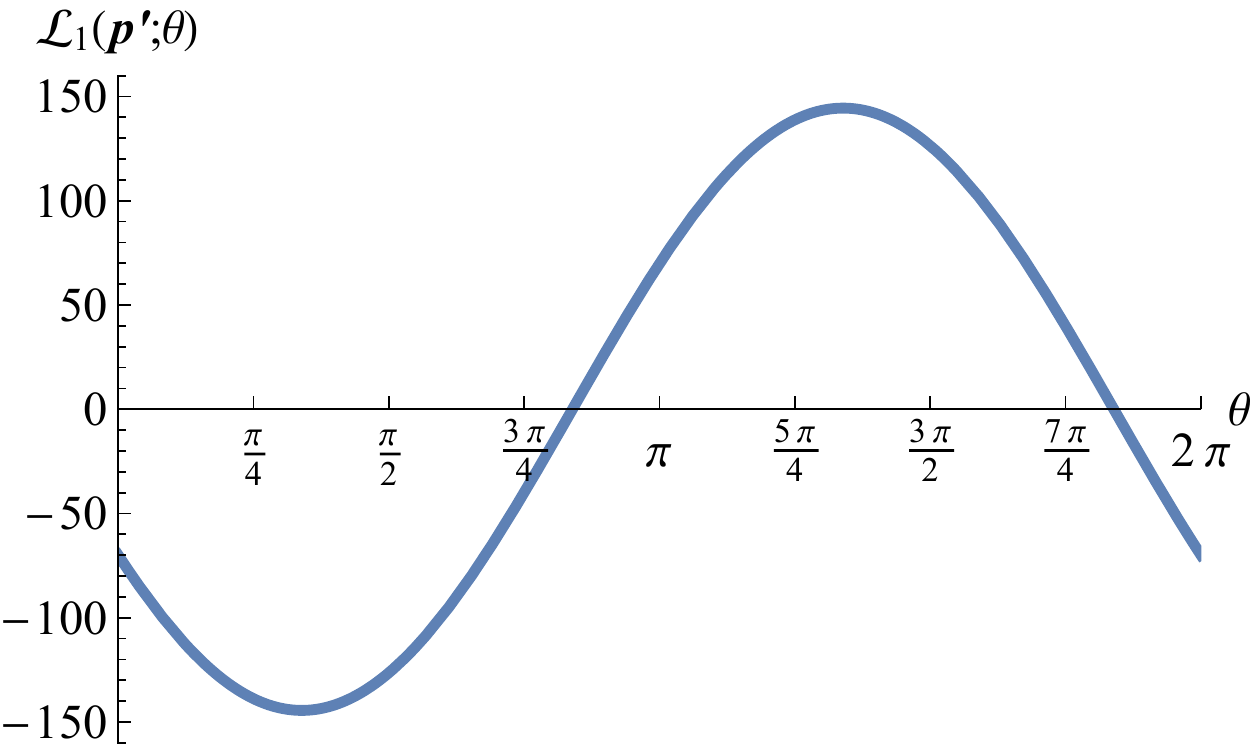}}
\end{minipage}
    \caption{ (a), (c), and (e): The fraction of protons in habitable dark-matter halos, as a function of our two-dimensional parameter space, for the cases $n=2/3, 2$, and $3$, respectively. In each case, the red dot corresponds to $f_{h}\sim 0.2$ and denotes the point in parameter space consistent with cosmological observables. More specifically, that dot denotes ${\bm p'}$ in the main text [see, e.g., Eq.~(\ref{EQN:Lform1})] and takes the values: ${\bm p'}=(-122.94, -11.72)$ for (a); ${\bm p'}=(-122.94, -12.16)$ for (c); and ${\bm p'}=(-122.94, -12.55)$ for (e). Levels of local fine-tuning at ${\bm p'}$ are displayed in (b), (d), and (f), as computed from (a), (c), and (e), respectively. The maximal level of local fine-tuning is of Order 145 in each case.}
        \label{FIG:LocalFTHH}
\end{figure*}

Our results for the observable, $f_{h}$, for the cases where $n=2/3, 2$, and $3$, are displayed in Fig.~\ref{FIG:LocalFTHH}(a), (c), and (e), respectively. In each of these plots, the red dot marks parameter values that are consistent with observations (as described above) --- and denotes the point, ${\bm p'}$, that enters into our computations of local and global fine-tuning [see, e.g., Eq.~(\ref{EQN:Lform1})]. In each case, we find, at the red dot, $f_{h}\sim 0.2$.

In computing levels of local fine-tuning via Eq.~(\ref{EQN:Lform1}) we again note that our parameter space is two-dimensional, so we can write $\hat{\bm{\epsilon}}=(\cos\theta, \sin\theta)$. Thus, once we have fixed some ${\bm p'}$, the level of local fine-tuning, $\mathcal{L}(\bm{p'};\theta)$ [or, alternatively, $\mathcal{L}_{1}(\bm{p'};\theta)$], is purely a function of the angle $\theta$. We display results for $\mathcal{L}_{1}(\bm{p'};\theta)$ for each of the three cases where $n=2/3,2$, and $3$, in Fig.~\ref{FIG:LocalFTHH}(b), (d), and (f), respectively, to obtain a maximal level of local fine-tuning of Order 145 in each case. The results across the three distinct cases are virtually identical. According to the classification scheme defined in Eqs.~(\ref{EQN:LocalClassif}), this corresponds to a {\it very significant} level of local fine-tuning. 

The computation of levels of global fine-tuning require a  parameter space where each parameter takes a finite range of possible values. In what follows, for the sake of illustration, we fix this range to correspond to $\rho_{\Lambda} \in [0.1 \rho_{\Lambda, {\rm obs}}, 1]$ and $\lambda \in [10^{-10}\lambda_{\rm obs}(n,N_{*}),1]$. In each case, the upper limits in the ranges reflect the fact that order-unity quantities in the units we employ are to be considered large. The lower limits in each case are largely illustrative.

\begin{figure*}
\begin{minipage}{.325\linewidth}
\centering
\subfloat[]{\includegraphics[scale=0.5]{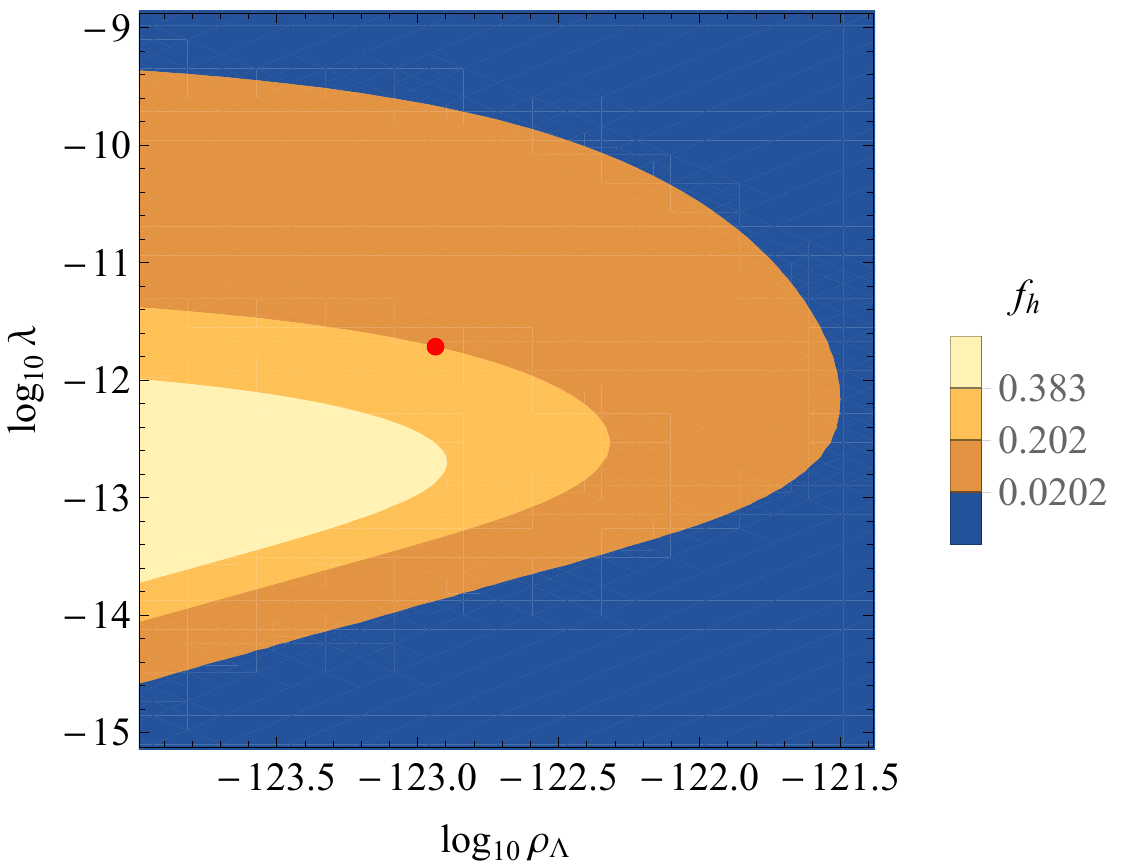}}
\end{minipage}
\begin{minipage}{.325\linewidth}
\centering
\subfloat[]{\includegraphics[scale=0.44]{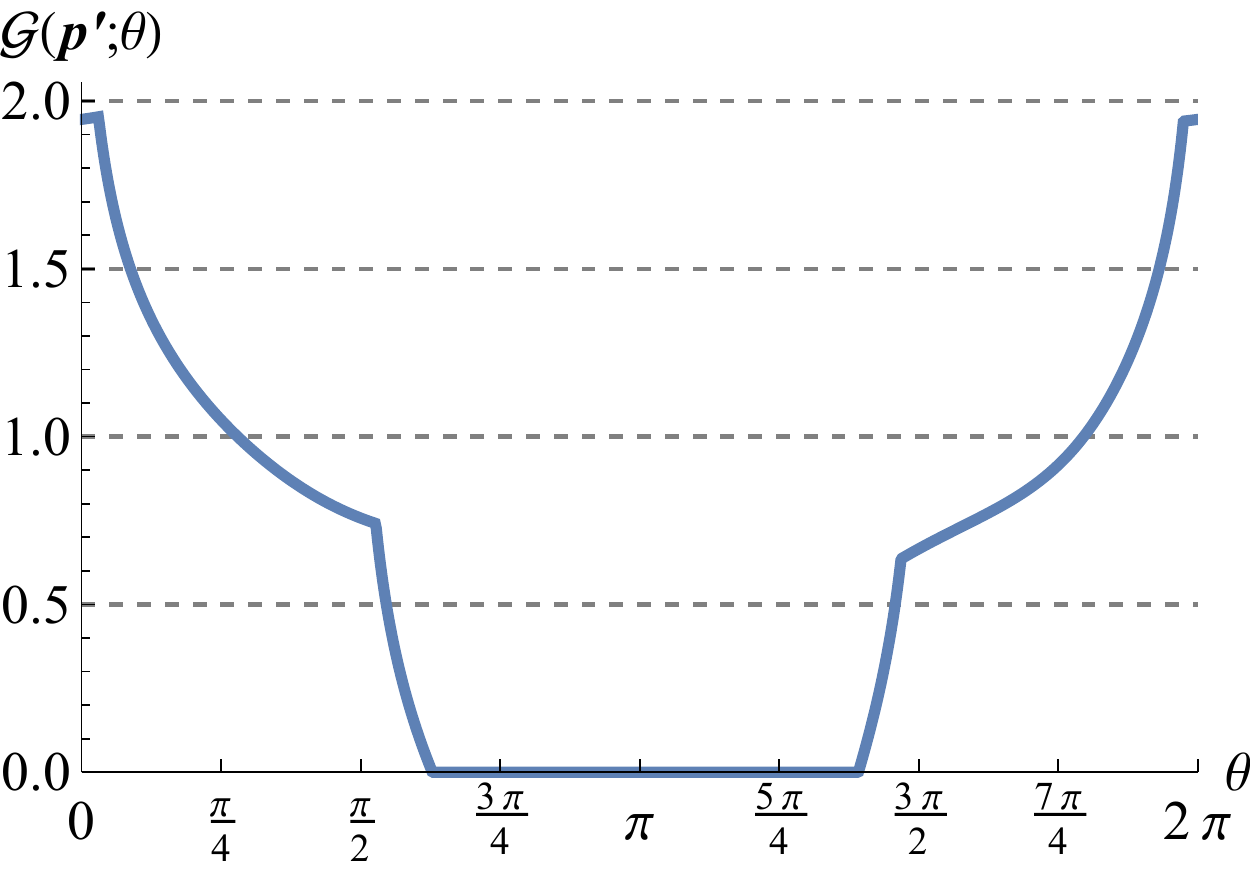}}
\end{minipage}
\begin{minipage}{.325\linewidth}
\centering
\subfloat[]{\includegraphics[scale=0.44]{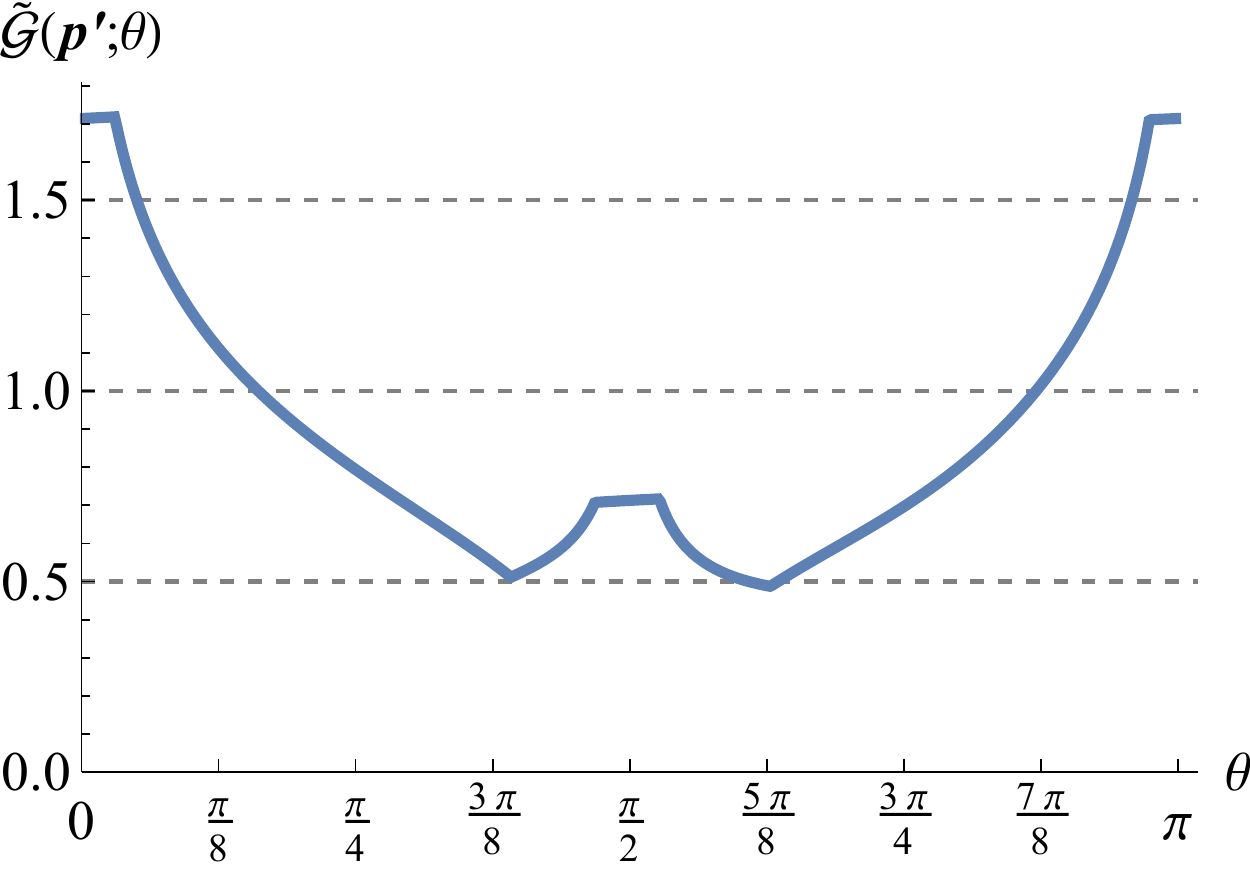}}
\end{minipage}
\begin{minipage}{.325\linewidth}
\centering
\subfloat[]{\includegraphics[scale=0.5]{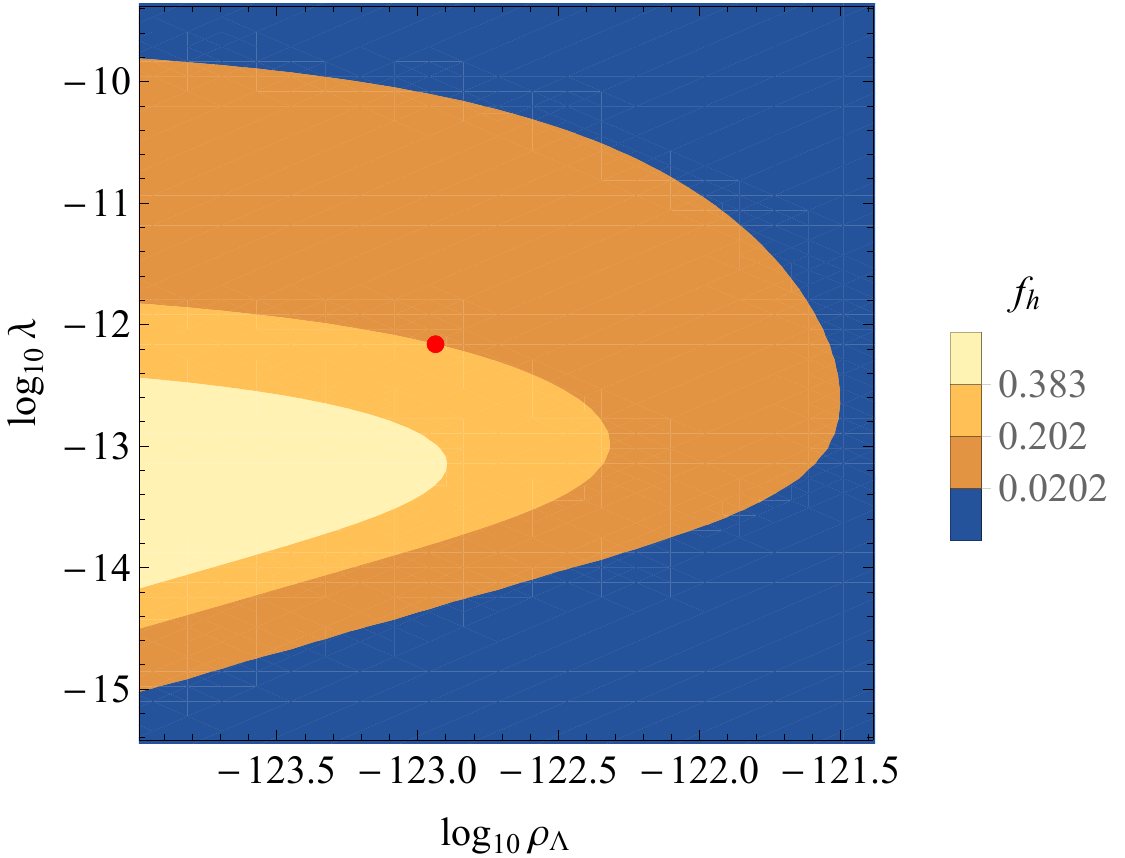}}
\end{minipage}
\begin{minipage}{.325\linewidth}
\centering
\subfloat[]{\includegraphics[scale=0.44]{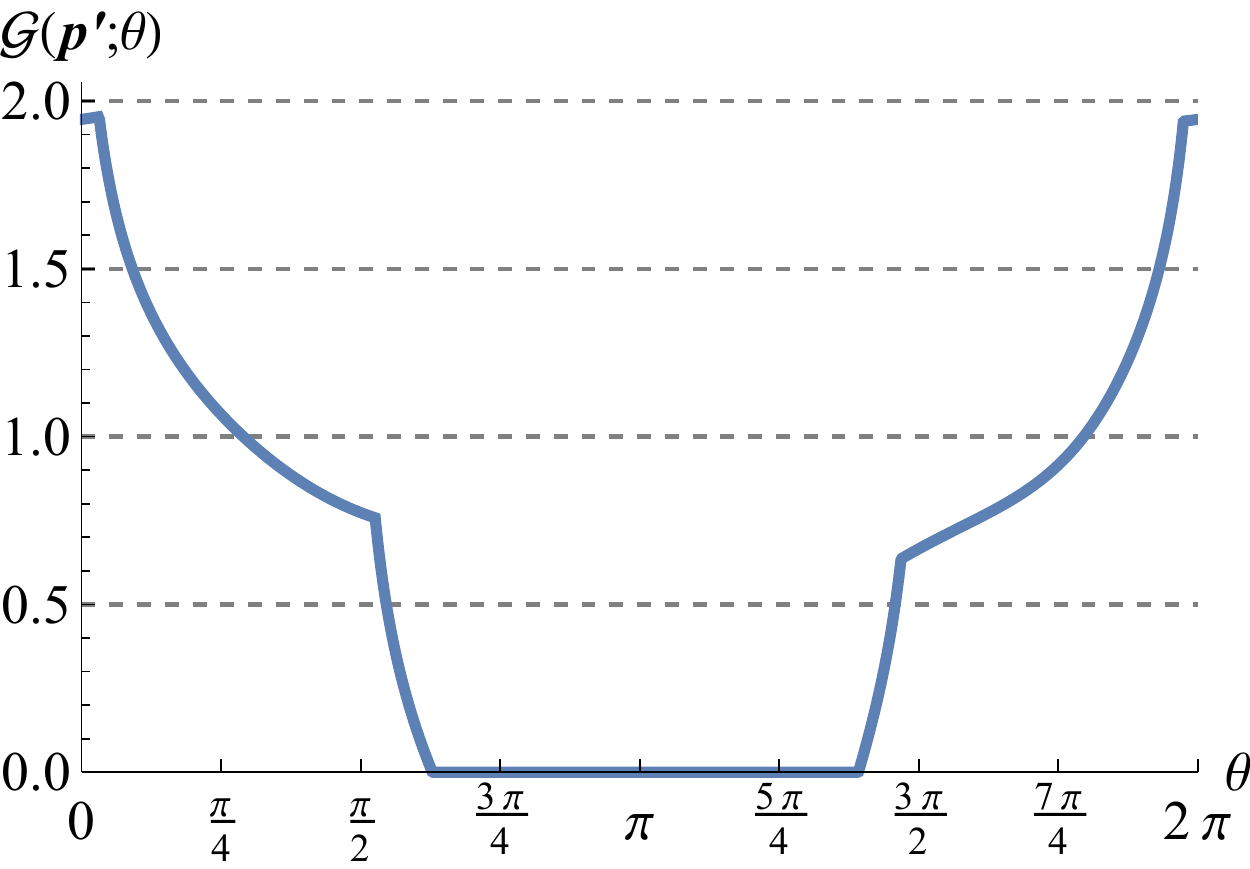}}
\end{minipage}
\begin{minipage}{.325\linewidth}
\centering
\subfloat[]{\includegraphics[scale=0.44]{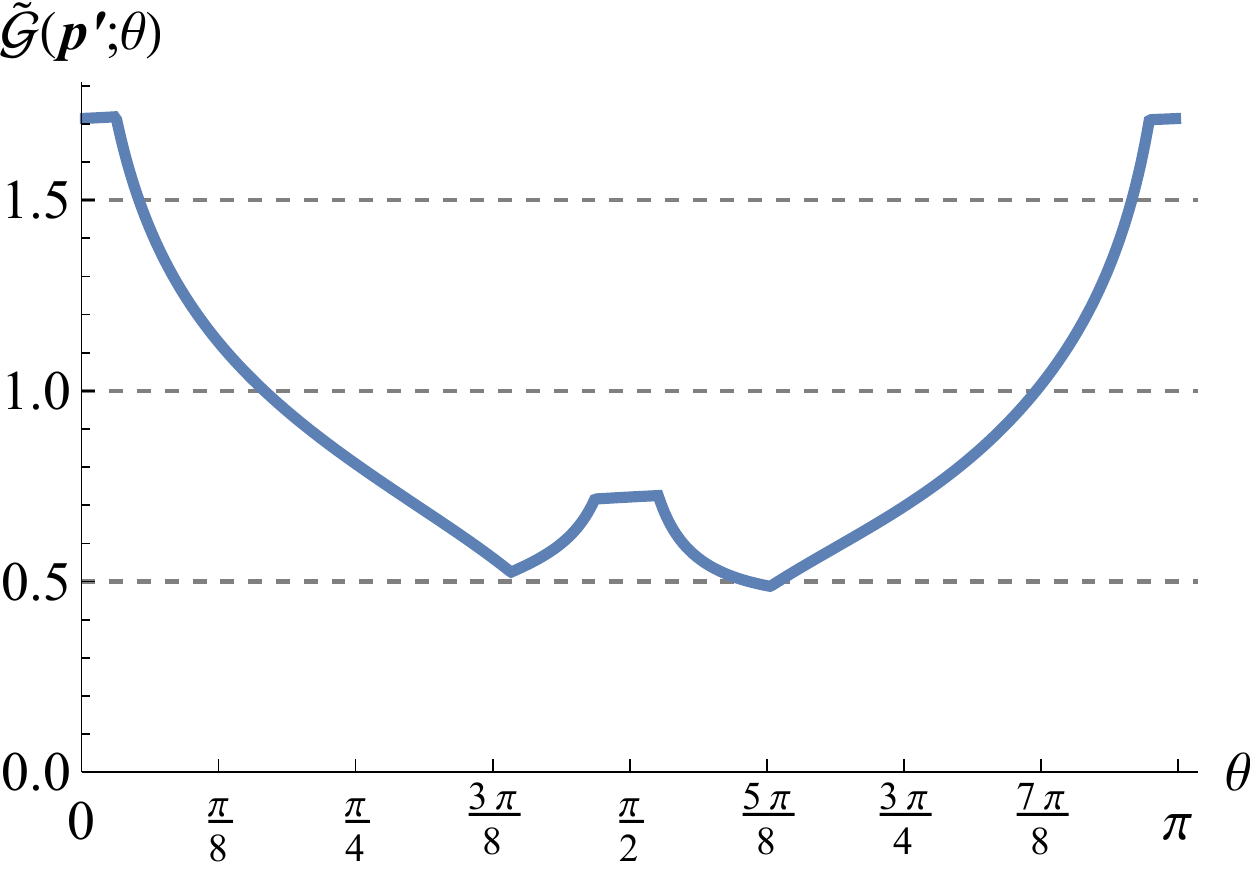}}
\end{minipage}
\begin{minipage}{.325\linewidth}
\centering
\subfloat[]{\includegraphics[scale=0.5]{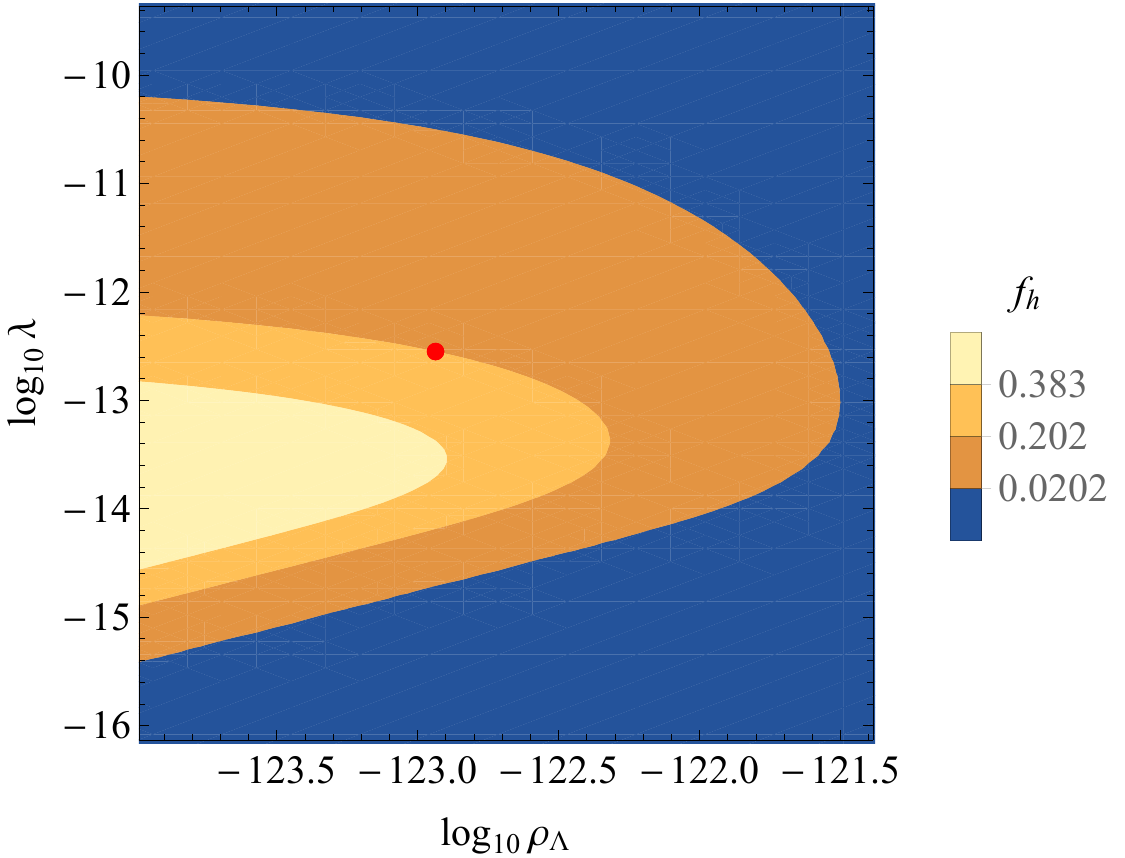}}
\end{minipage}
\begin{minipage}{.325\linewidth}
\centering
\subfloat[]{\includegraphics[scale=0.44]{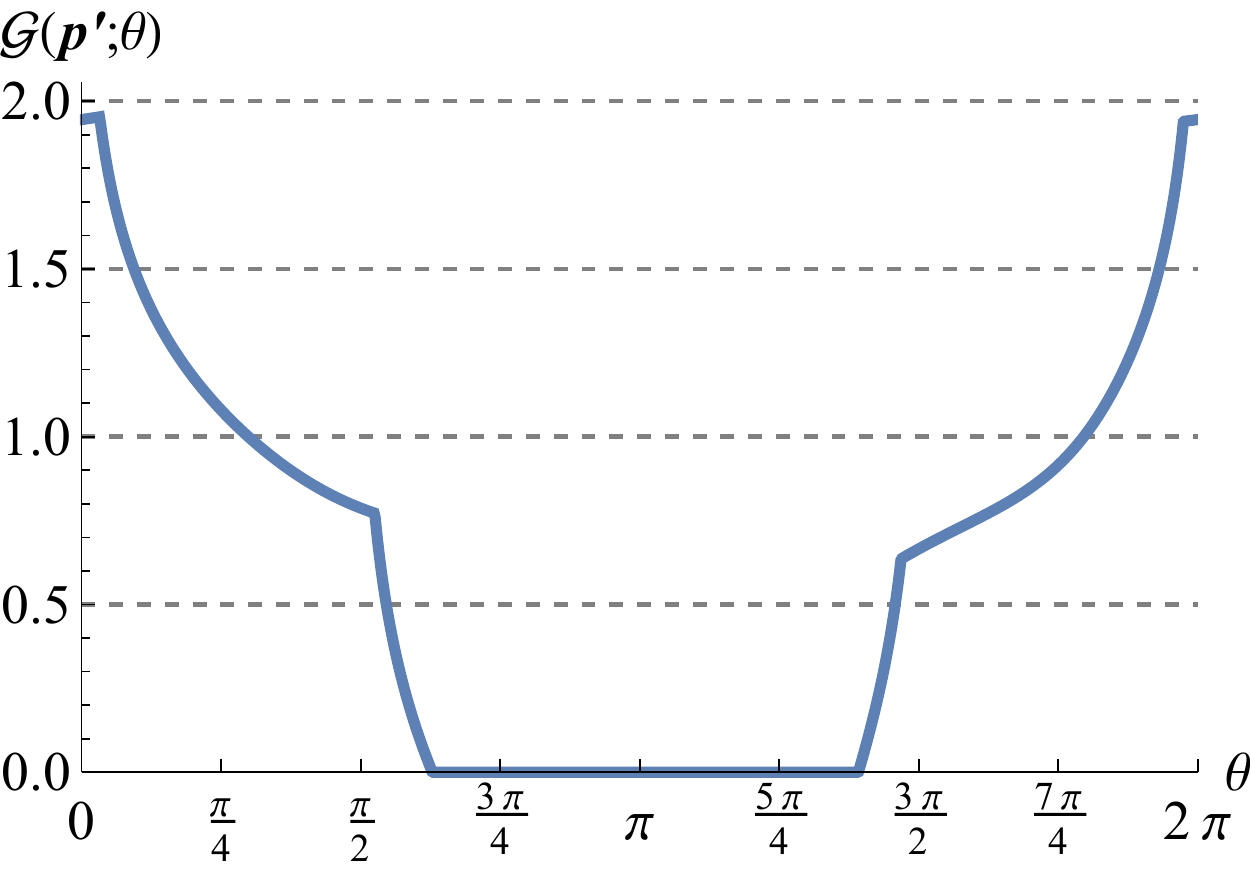}}
\end{minipage}
\begin{minipage}{.325\linewidth}
\centering
\subfloat[]{\includegraphics[scale=0.44]{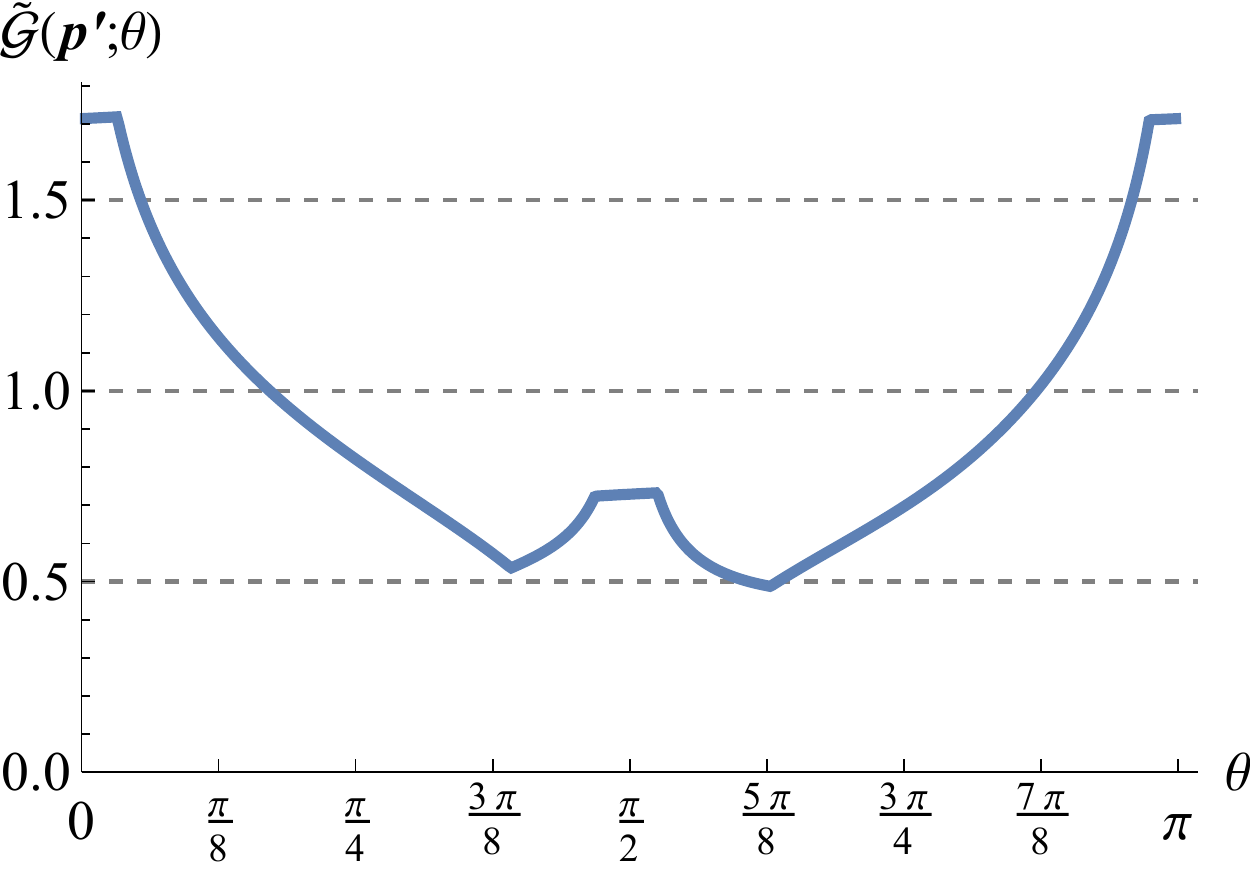}}
\end{minipage}
    \caption{Levels of global fine-tuning of  $f_{h}$, the fraction of protons in habitable dark-matter halos as a function of inflationary parameters [appearing in Eq.~(\ref{EQN:Vphi})]. (a)--(c) display results for $n=2/3$; (d)--(f) display results for $n=2$; (g)--(i) display results for $n=3$. (a), (d), and (g) present a close-up of the corresponding region in Fig.~\ref{FIG:LocalFTHH}(a), (c), and (e), respectively. Contours mark order-unity increases and decreases in $f_{h}$, though only order-unity decreases are used in our global analysis. A red dot corresponds to the point in parameter space consistent with measured cosmological observables, as quoted in the caption to Fig.~\ref{FIG:LocalFTHH}. Subplots (b), (e), and (h) present computations of the measure of global fine-tuning in Eq.~(\ref{EQN:Global}), for $n=2/3, 2$ and $3$, respectively. The dashed gray lines demarcate boundaries where the order of global fine-tuning changes [see Eqs.~(\ref{EQN:GlobalOrdDEF})]. As expected, $\mathcal{G}({\bm p';\theta})=0$ in directions that reach the edge of parameter space before an order-unity decrease in $f_{h}$ occurs. (c), (f), and (i) present computations of the measure of global fine-tuning in Eq.~(\ref{EQN:Global2}).}\label{FIG:GlobalFTHH}
\end{figure*}

Measures of global fine-tuning can be computed using Eqs.~(\ref{EQN:OrderUNITY})--(\ref{EQN:Global2}). Directions in parameter space (labeled by, for example, $\hat{\bm v}$ in Sec.~\ref{SEC:GlobalMeasure}) can again be parameterized by $\theta$. For each direction $\theta$ we may now use Eq.~(\ref{EQN:OrderUNITY}) to find the length of the vector that yields an order-unity change in the observable, $f_h$. Since, as described in Sec.~\ref{SEC:GlobalMeasure}, we are interested in the potential for dark-matter halos to give rise to life, an order-unity increase in $f_{h}$ is not significant, but an order-unity decrease is. We choose the order-unity decrease to correspond to $0.9$ [viz., the right-hand side of Eq.~(\ref{EQN:OrderUNITY}) is set to 0.9 and only a decrease in the value of the observable is deemed significant], and display numerical results for our computation of $\mathcal{G}({\bm p'};\theta)$ and $\tilde{\mathcal{G}}({\bm p'};\theta)$, for each of the three cases $n=2/3,2$, and $3$, in Fig.~\ref{FIG:GlobalFTHH}.

In Fig.~\ref{FIG:GlobalFTHH}(b), (e), and (h), we see that the maximal level of global fine-tuning is of Order 4 [i.e., when $3/2 < \mathcal{G}({\bm p';\theta})\leq 2$ --- according to Eqs.~(\ref{EQN:GlobalOrdDEF})], and that $\mathcal{G}({\bm p';\theta})=0$ when the direction of interest does not yield a significant order-unity change in the observable before one reaches the edge of parameter space. When we take into account the entire range of parameter values along a line through ${\bm p'}$, as in the computation of $\tilde{\mathcal{G}}({\bm p';\theta})$, we find that the minimum level of global fine-tuning is of Order 1 for each value of $n$ considered. Again, results for each case are very similar. In sum: the fraction of protons in habitable dark-matter halos, consistent with observations, can require extreme levels of local fine-tuning and high levels of global fine-tuning for the models of cosmic inflation described above.

\section{Discussion}\label{SEC:Discussion}

In this paper, we have defined quantitative measures of fine-tuning from `local' and `global' perspectives, suited to the assessment of various theories (or models) employed in the sciences. The underlying motivation for this work is the claim that the perceived level of fine-tuning of scientific theories plays an important role in the evaluation and development of theories (toward less-finely tuned theories) --- and yet it remains a difficult challenge to make precise what we mean by fine-tuning. 

There are many examples in the history of science that may be interpreted to support the underlying motivation mentioned above. One example comes from early advances in astronomy, in which the geocentric model of Ptolemy (indeed, described by~\citet{weinberg_15} as finely tuned) was supplanted by Copernicus' heliocentric model. Another example, not related to physics per se, arises in {\it On the Origin of Species}, where Darwin argued for the hypothesis of common ancestry over (what we would naturally describe as) the finely tuned hypothesis of special creation~\cite{lewens_07}: with the former hypothesis furnishing a better explanation for, say, the existence of similar features of inhabitants of regions with ostensibly different climatic conditions. And, skipping ahead to the present day, it is interesting (and perhaps unsurprising) to note that those in the vanguard of theory development in emerging disciplines such as in theoretical biophysics are indeed sensitive to such a tradition (see, for overviews, Refs.~\citep{bialek_12, bialek_15}, and references therein).

The groundwork we have laid for meeting the challenge of making fine-tuning precise, describes local and global measures suited to the analysis of a broad range of such scientific settings. Our measures can indeed be employed to establish and compare levels of fine-tuning in various contexts. In an example of an application of our formalism, in Sec.~\ref{SEC:PBHs}, we characterized fine-tuning in models where a significant fraction of dark matter is in the form of PBHs, and found these scenarios to be finely tuned in both a local and global sense. We expect this fine-tuning to increase substantially if one traces the evolution of PBHs from, for example, inflationary mechanisms up to the generation of extended mass functions.

In a second example, described in Sec.~\ref{SEC:HabitableHalos}, which indeed involves inflationary processes in the very early universe, we characterized fine-tuning in models that derive the fraction of protons in habitable dark-matter halos from underlying models of cosmic inflation. We found these scenarios to be significantly finely tuned in both a local and global sense, with levels of fine-tuning being very similar across the three inflationary models studied. 

We also wish to highlight some important caveats to our approach. In referring to sizes of ranges in parameter space, our global measure of fine-tuning implicitly treats different points in parameter space equally. If one wishes to assign different weights to different points in parameter space, a new global measure of fine-tuning would need to be constructed. Indeed, we have not dealt with the important issue of measure-theoretic descriptions of fine-tuning, but they are the natural setting in which to describe such a generalization of our scheme. Of course, the central feature of our definition of global fine-tuning would remain in place. That is, establishing levels of global fine-tuning consists of a quantitative comparison between (i) a measure of all relevant parameter values, and (ii) the (smaller) measure of such parameter values that are consistent with the phenomenon whose finely tuned nature is under investigation. 

Finally, we note that fine-tuning is just one aspect of a theory (or model) that may be used in its assessment. There are other characteristics that may also be taken into account when evaluating and comparing theories, such as empirical adequacy, internal and external consistency, scope, simplicity, and the ability of a theory to provoke new lines of inquiry. (See, for example, Refs.~\cite{kuhn_77, mcmullin_83}.) Thus we caution against using {\it only} the measures we have explicated in this paper (or more generalized versions of such measures) to assess a theory, but stress that such measures are an important characteristic of theories, and are a valuable aid in the difficult task of theory development considered broadly across the sciences.

\section*{Acknowledgements} 
F.~A.~thanks Jeremy Butterfield and Jos Uffink for discussions. We acknowledge support from the Black Hole Initiative at Harvard University, which is funded through a grant from the John Templeton Foundation.

\bibliography{azhar+loeb_18.bib}

\begin{thebibliography}{56}%
\makeatletter
\providecommand \@ifxundefined [1]{%
 \@ifx{#1\undefined}
}%
\providecommand \@ifnum [1]{%
 \ifnum #1\expandafter \@firstoftwo
 \else \expandafter \@secondoftwo
 \fi
}%
\providecommand \@ifx [1]{%
 \ifx #1\expandafter \@firstoftwo
 \else \expandafter \@secondoftwo
 \fi
}%
\providecommand \natexlab [1]{#1}%
\providecommand \enquote  [1]{``#1''}%
\providecommand \bibnamefont  [1]{#1}%
\providecommand \bibfnamefont [1]{#1}%
\providecommand \citenamefont [1]{#1}%
\providecommand \href@noop [0]{\@secondoftwo}%
\providecommand \href [0]{\begingroup \@sanitize@url \@href}%
\providecommand \@href[1]{\@@startlink{#1}\@@href}%
\providecommand \@@href[1]{\endgroup#1\@@endlink}%
\providecommand \@sanitize@url [0]{\catcode `\\12\catcode `\$12\catcode
  `\&12\catcode `\#12\catcode `\^12\catcode `\_12\catcode `\%12\relax}%
\providecommand \@@startlink[1]{}%
\providecommand \@@endlink[0]{}%
\providecommand \url  [0]{\begingroup\@sanitize@url \@url }%
\providecommand \@url [1]{\endgroup\@href {#1}{\urlprefix }}%
\providecommand \urlprefix  [0]{URL }%
\providecommand \Eprint [0]{\href }%
\providecommand \doibase [0]{http://dx.doi.org/}%
\providecommand \selectlanguage [0]{\@gobble}%
\providecommand \bibinfo  [0]{\@secondoftwo}%
\providecommand \bibfield  [0]{\@secondoftwo}%
\providecommand \translation [1]{[#1]}%
\providecommand \BibitemOpen [0]{}%
\providecommand \bibitemStop [0]{}%
\providecommand \bibitemNoStop [0]{.\EOS\space}%
\providecommand \EOS [0]{\spacefactor3000\relax}%
\providecommand \BibitemShut  [1]{\csname bibitem#1\endcsname}%
\let\auto@bib@innerbib\@empty
%</preamble>
\bibitem [{\citenamefont {{Carter}}(1974)}]{carter_74}%
  \BibitemOpen
  \bibfield  {author} {\bibinfo {author} {\bibfnamefont {B.}~\bibnamefont
  {{Carter}}},\ }\bibfield  {title} {\enquote {\bibinfo {title} {{Large number
  coincidences and the anthropic principle in cosmology}},}\ }in\ \href@noop {}
  {\emph {\bibinfo {booktitle} {Confrontation of Cosmological Theories with
  Observational Data}}},\ \bibinfo {series} {IAU Symposium}, Vol.~\bibinfo
  {volume} {63},\ \bibinfo {editor} {edited by\ \bibinfo {editor}
  {\bibfnamefont {M.~S.}\ \bibnamefont {{Longair}}}}\ (\bibinfo {year} {1974})\
  p.\ \bibinfo {pages} {291}\BibitemShut {NoStop}%
\bibitem [{\citenamefont {{Carr}}\ and\ \citenamefont
  {{Rees}}(1979)}]{carr+rees_79}%
  \BibitemOpen
  \bibfield  {author} {\bibinfo {author} {\bibfnamefont {B.~J.}\ \bibnamefont
  {{Carr}}}\ and\ \bibinfo {author} {\bibfnamefont {M.~J.}\ \bibnamefont
  {{Rees}}},\ }\bibfield  {title} {\enquote {\bibinfo {title} {The anthropic
  principle and the structure of the physical world},}\ }\href
  {https://doi.org/10.1038/278605a0} {\bibfield  {journal} {\bibinfo  {journal}
  {\nat}\ }\textbf {\bibinfo {volume} {278}},\ \bibinfo {pages} {605} (\bibinfo
  {year} {1979})}\BibitemShut {NoStop}%
\bibitem [{\citenamefont {Barrow}\ and\ \citenamefont
  {Tipler}(1986)}]{barrow+tipler_86}%
  \BibitemOpen
  \bibfield  {author} {\bibinfo {author} {\bibfnamefont {J.~D.}\ \bibnamefont
  {Barrow}}\ and\ \bibinfo {author} {\bibfnamefont {F.~J.}\ \bibnamefont
  {Tipler}},\ }\href@noop {} {\emph {\bibinfo {title} {The {A}nthropic
  {C}osmological {P}rinciple}}}\ (\bibinfo  {publisher} {Oxford University
  Press, Oxford, UK},\ \bibinfo {year} {1986})\BibitemShut {NoStop}%
\bibitem [{\citenamefont {{Tegmark}}\ \emph {et~al.}(2006)\citenamefont
  {{Tegmark}}, \citenamefont {{Aguirre}}, \citenamefont {{Rees}},\ and\
  \citenamefont {{Wilczek}}}]{tegmark+al_06}%
  \BibitemOpen
  \bibfield  {author} {\bibinfo {author} {\bibfnamefont {M.}~\bibnamefont
  {{Tegmark}}}, \bibinfo {author} {\bibfnamefont {A.}~\bibnamefont
  {{Aguirre}}}, \bibinfo {author} {\bibfnamefont {M.~J.}\ \bibnamefont
  {{Rees}}}, \ and\ \bibinfo {author} {\bibfnamefont {F.}~\bibnamefont
  {{Wilczek}}},\ }\bibfield  {title} {\enquote {\bibinfo {title}
  {{Dimensionless constants, cosmology, and other dark matters}},}\ }\href
  {https://doi.org/10.1103/PhysRevD.73.023505} {\bibfield  {journal} {\bibinfo
  {journal} {\prd}\ }\textbf {\bibinfo {volume} {73}},\ \bibinfo {eid} {023505}
  (\bibinfo {year} {2006})},\ \Eprint {http://arxiv.org/abs/astro-ph/0511774}
  {astro-ph/0511774} \BibitemShut {NoStop}%
\bibitem [{\citenamefont {Maudlin}(2017)}]{maudlin_crete}%
  \BibitemOpen
  \bibfield  {author} {\bibinfo {author} {\bibfnamefont {T.}~\bibnamefont
  {Maudlin}},\ }\bibfield  {title} {\enquote {\bibinfo {title} {Fine-tuned for
  what?}}\ }\href
  {https://finetune.physics.ox.ac.uk/resources/international-conference-physics-fine-tuning}
  {\bibfield  {journal} {\bibinfo  {journal} {International Conference on the
  Physics of Fine-Tuning}\ } (\bibinfo {year} {2017})},\ \bibinfo {note}
  {accessed July 7, 2018}\BibitemShut {NoStop}%
\bibitem [{\citenamefont {{Guth}}(1981)}]{guth_81}%
  \BibitemOpen
  \bibfield  {author} {\bibinfo {author} {\bibfnamefont {A.~H.}\ \bibnamefont
  {{Guth}}},\ }\bibfield  {title} {\enquote {\bibinfo {title} {{Inflationary
  universe: A possible solution to the horizon and flatness problems}},}\
  }\href {https://doi.org/10.1103/PhysRevD.23.347} {\bibfield  {journal}
  {\bibinfo  {journal} {\prd}\ }\textbf {\bibinfo {volume} {23}},\ \bibinfo
  {pages} {347} (\bibinfo {year} {1981})}\BibitemShut {NoStop}%
\bibitem [{\citenamefont {{Linde}}(1982)}]{linde_82}%
  \BibitemOpen
  \bibfield  {author} {\bibinfo {author} {\bibfnamefont {A.~D.}\ \bibnamefont
  {{Linde}}},\ }\bibfield  {title} {\enquote {\bibinfo {title} {{A new
  inflationary universe scenario: A possible solution of the horizon, flatness,
  homogeneity, isotropy and primordial monopole problems}},}\ }\href
  {https://doi.org/10.1016/0370-2693(82)91219-9} {\bibfield  {journal}
  {\bibinfo  {journal} {Phys. Lett. B}\ }\textbf {\bibinfo {volume} {108}},\
  \bibinfo {pages} {389} (\bibinfo {year} {1982})}\BibitemShut {NoStop}%
\bibitem [{\citenamefont {{Albrecht}}\ and\ \citenamefont
  {{Steinhardt}}(1982)}]{albrecht+steinhardt_82}%
  \BibitemOpen
  \bibfield  {author} {\bibinfo {author} {\bibfnamefont {A.}~\bibnamefont
  {{Albrecht}}}\ and\ \bibinfo {author} {\bibfnamefont {P.~J.}\ \bibnamefont
  {{Steinhardt}}},\ }\bibfield  {title} {\enquote {\bibinfo {title} {{Cosmology
  for grand unified theories with radiatively induced symmetry breaking}},}\
  }\href {https://doi.org/10.1103/PhysRevLett.48.1220} {\bibfield  {journal}
  {\bibinfo  {journal} {Phys. Rev. Lett.}\ }\textbf {\bibinfo {volume} {48}},\
  \bibinfo {pages} {1220} (\bibinfo {year} {1982})}\BibitemShut {NoStop}%
\bibitem [{\citenamefont {{Linde}}(1983)}]{linde_83b}%
  \BibitemOpen
  \bibfield  {author} {\bibinfo {author} {\bibfnamefont {A.~D.}\ \bibnamefont
  {{Linde}}},\ }\bibfield  {title} {\enquote {\bibinfo {title} {{Chaotic
  inflation}},}\ }\href {https://doi.org/10.1016/0370-2693(83)90837-7}
  {\bibfield  {journal} {\bibinfo  {journal} {Phys. Lett. B}\ }\textbf
  {\bibinfo {volume} {129}},\ \bibinfo {pages} {177} (\bibinfo {year}
  {1983})}\BibitemShut {NoStop}%
\bibitem [{\citenamefont {{Ijjas}}\ \emph {et~al.}(2013)\citenamefont
  {{Ijjas}}, \citenamefont {{Steinhardt}},\ and\ \citenamefont
  {{Loeb}}}]{ijjas+al_13}%
  \BibitemOpen
  \bibfield  {author} {\bibinfo {author} {\bibfnamefont {A.}~\bibnamefont
  {{Ijjas}}}, \bibinfo {author} {\bibfnamefont {P.~J.}\ \bibnamefont
  {{Steinhardt}}}, \ and\ \bibinfo {author} {\bibfnamefont {A.}~\bibnamefont
  {{Loeb}}},\ }\bibfield  {title} {\enquote {\bibinfo {title} {{Inflationary
  paradigm in trouble after Planck2013}},}\ }\href
  {https://doi.org/10.1016/j.physletb.2013.05.023} {\bibfield  {journal}
  {\bibinfo  {journal} {Phys. Lett. B}\ }\textbf {\bibinfo {volume} {723}},\
  \bibinfo {pages} {261} (\bibinfo {year} {2013})},\ \Eprint
  {http://arxiv.org/abs/1304.2785} {arXiv:1304.2785} \BibitemShut {NoStop}%
\bibitem [{\citenamefont {{Guth}}\ \emph {et~al.}(2014)\citenamefont {{Guth}},
  \citenamefont {{Kaiser}},\ and\ \citenamefont {{Nomura}}}]{guth+al_14}%
  \BibitemOpen
  \bibfield  {author} {\bibinfo {author} {\bibfnamefont {A.~H.}\ \bibnamefont
  {{Guth}}}, \bibinfo {author} {\bibfnamefont {D.~I.}\ \bibnamefont
  {{Kaiser}}}, \ and\ \bibinfo {author} {\bibfnamefont {Y.}~\bibnamefont
  {{Nomura}}},\ }\bibfield  {title} {\enquote {\bibinfo {title} {{Inflationary
  paradigm after Planck 2013}},}\ }\href
  {https://doi.org/10.1016/j.physletb.2014.03.020} {\bibfield  {journal}
  {\bibinfo  {journal} {Phys. Lett. B}\ }\textbf {\bibinfo {volume} {733}},\
  \bibinfo {pages} {112} (\bibinfo {year} {2014})},\ \Eprint
  {http://arxiv.org/abs/1312.7619} {arXiv:1312.7619} \BibitemShut {NoStop}%
\bibitem [{\citenamefont {Ellis}\ \emph {et~al.}(1986)\citenamefont {Ellis},
  \citenamefont {Enqvist}, \citenamefont {Nanopoulos},\ and\ \citenamefont
  {Zwirner}}]{ellis+al_86}%
  \BibitemOpen
  \bibfield  {author} {\bibinfo {author} {\bibfnamefont {J.}~\bibnamefont
  {Ellis}}, \bibinfo {author} {\bibfnamefont {K.}~\bibnamefont {Enqvist}},
  \bibinfo {author} {\bibfnamefont {D.~V.}\ \bibnamefont {Nanopoulos}}, \ and\
  \bibinfo {author} {\bibfnamefont {F.}~\bibnamefont {Zwirner}},\ }\bibfield
  {title} {\enquote {\bibinfo {title} {Observables in low-energy superstring
  models},}\ }\href {https://doi.org/10.1142/S0217732386000105} {\bibfield
  {journal} {\bibinfo  {journal} {Mod. Phys. Lett. A}\ }\textbf {\bibinfo
  {volume} {1}},\ \bibinfo {pages} {57} (\bibinfo {year} {1986})}\BibitemShut
  {NoStop}%
\bibitem [{\citenamefont {Barbieri}\ and\ \citenamefont
  {Giudice}(1988)}]{barbieri+giudice_88}%
  \BibitemOpen
  \bibfield  {author} {\bibinfo {author} {\bibfnamefont {R.}~\bibnamefont
  {Barbieri}}\ and\ \bibinfo {author} {\bibfnamefont {G.~F.}\ \bibnamefont
  {Giudice}},\ }\bibfield  {title} {\enquote {\bibinfo {title} {Upper bounds on
  supersymmetric particle masses},}\ }\href
  {https://doi.org/10.1016/0550-3213(88)90171-X} {\bibfield  {journal}
  {\bibinfo  {journal} {Nucl. Phys. B}\ }\textbf {\bibinfo {volume} {306}},\
  \bibinfo {pages} {63} (\bibinfo {year} {1988})}\BibitemShut {NoStop}%
\bibitem [{\citenamefont {Anderson}\ and\ \citenamefont
  {Casta\~{n}o}(1995)}]{anderson+castano_95}%
  \BibitemOpen
  \bibfield  {author} {\bibinfo {author} {\bibfnamefont {G.~W.}\ \bibnamefont
  {Anderson}}\ and\ \bibinfo {author} {\bibfnamefont {D.~J.}\ \bibnamefont
  {Casta\~{n}o}},\ }\bibfield  {title} {\enquote {\bibinfo {title} {{Measures
  of fine tuning}},}\ }\href {https://doi.org/10.1016/0370-2693(95)00051-L}
  {\bibfield  {journal} {\bibinfo  {journal} {Phys. Lett. B}\ }\textbf
  {\bibinfo {volume} {347}},\ \bibinfo {pages} {300} (\bibinfo {year}
  {1995})},\ \Eprint {http://arxiv.org/abs/hep-ph/9409419}
  {arXiv:hep-ph/9409419} \BibitemShut {NoStop}%
\bibitem [{\citenamefont {Athron}\ and\ \citenamefont
  {Miller}(2007)}]{athron+miller_07}%
  \BibitemOpen
  \bibfield  {author} {\bibinfo {author} {\bibfnamefont {P.}~\bibnamefont
  {Athron}}\ and\ \bibinfo {author} {\bibfnamefont {D.~J.}\ \bibnamefont
  {Miller}},\ }\bibfield  {title} {\enquote {\bibinfo {title} {New measure of
  fine tuning},}\ }\href {https://doi.org/10.1103/PhysRevD.76.075010}
  {\bibfield  {journal} {\bibinfo  {journal} {Phys. Rev. D}\ }\textbf {\bibinfo
  {volume} {76}},\ \bibinfo {pages} {075010} (\bibinfo {year} {2007})},\
  \Eprint {http://arxiv.org/abs/0705.2241} {arXiv:0705.2241} \BibitemShut
  {NoStop}%
\bibitem [{\citenamefont {{Barnes}}(2018)}]{barnes_17}%
  \BibitemOpen
  \bibfield  {author} {\bibinfo {author} {\bibfnamefont {L.~A.}\ \bibnamefont
  {{Barnes}}},\ }\bibfield  {title} {\enquote {\bibinfo {title} {Fine-tuning in
  the context of {B}ayesian theory testing},}\ }\href
  {https://doi.org/10.1007/s13194-017-0184-2} {\bibfield  {journal} {\bibinfo
  {journal} {Euro. J. Phil. Sci.}\ }\textbf {\bibinfo {volume} {8}},\ \bibinfo
  {pages} {253} (\bibinfo {year} {2018})},\ \Eprint
  {http://arxiv.org/abs/1707.03965} {arXiv:1707.03965} \BibitemShut {NoStop}%
\bibitem [{\citenamefont {{Zel'dovich}}\ and\ \citenamefont
  {{Novikov}}(1967)}]{zeldovich+novikov_67}%
  \BibitemOpen
  \bibfield  {author} {\bibinfo {author} {\bibfnamefont {Y.~B.}\ \bibnamefont
  {{Zel'dovich}}}\ and\ \bibinfo {author} {\bibfnamefont {I.~D.}\ \bibnamefont
  {{Novikov}}},\ }\bibfield  {title} {\enquote {\bibinfo {title} {The
  hypothesis of cores retarded during expansion and the hot cosmological
  model},}\ }\href@noop {} {\bibfield  {journal} {\bibinfo  {journal} {Sov.
  Astron.}\ }\textbf {\bibinfo {volume} {10}},\ \bibinfo {pages} {602}
  (\bibinfo {year} {1967})}\BibitemShut {NoStop}%
\bibitem [{\citenamefont {{Hawking}}(1971)}]{hawking_71}%
  \BibitemOpen
  \bibfield  {author} {\bibinfo {author} {\bibfnamefont {S.}~\bibnamefont
  {{Hawking}}},\ }\bibfield  {title} {\enquote {\bibinfo {title}
  {Gravitationally collapsed objects of very low mass},}\ }\href
  {https://doi.org/10.1093/mnras/152.1.75} {\bibfield  {journal} {\bibinfo
  {journal} {Mon. Not. R. Astro. Soc.}\ }\textbf {\bibinfo {volume} {152}},\
  \bibinfo {pages} {75} (\bibinfo {year} {1971})}\BibitemShut {NoStop}%
\bibitem [{\citenamefont {{Carr}}\ and\ \citenamefont
  {{Hawking}}(1974)}]{carr+hawking_74}%
  \BibitemOpen
  \bibfield  {author} {\bibinfo {author} {\bibfnamefont {B.~J.}\ \bibnamefont
  {{Carr}}}\ and\ \bibinfo {author} {\bibfnamefont {S.~W.}\ \bibnamefont
  {{Hawking}}},\ }\bibfield  {title} {\enquote {\bibinfo {title} {{Black holes
  in the early Universe}},}\ }\href {https://doi.org/10.1093/mnras/168.2.399}
  {\bibfield  {journal} {\bibinfo  {journal} {Mon. Not. R. Astro. Soc.}\
  }\textbf {\bibinfo {volume} {168}},\ \bibinfo {pages} {399} (\bibinfo {year}
  {1974})}\BibitemShut {NoStop}%
\bibitem [{\citenamefont {{Carr}}(1975)}]{carr_75}%
  \BibitemOpen
  \bibfield  {author} {\bibinfo {author} {\bibfnamefont {B.~J.}\ \bibnamefont
  {{Carr}}},\ }\bibfield  {title} {\enquote {\bibinfo {title} {The primordial
  black hole mass spectrum},}\ }\href {https://doi.org/10.1086/153853}
  {\bibfield  {journal} {\bibinfo  {journal} {Astrophy. J.}\ }\textbf {\bibinfo
  {volume} {201}},\ \bibinfo {pages} {1} (\bibinfo {year} {1975})}\BibitemShut
  {NoStop}%
\bibitem [{\citenamefont {{Garc{\'{\i}}a-Bellido}}\ \emph
  {et~al.}(1996)\citenamefont {{Garc{\'{\i}}a-Bellido}}, \citenamefont
  {{Linde}},\ and\ \citenamefont {{Wands}}}]{garcia-bellido+al_96}%
  \BibitemOpen
  \bibfield  {author} {\bibinfo {author} {\bibfnamefont {J.}~\bibnamefont
  {{Garc{\'{\i}}a-Bellido}}}, \bibinfo {author} {\bibfnamefont
  {A.}~\bibnamefont {{Linde}}}, \ and\ \bibinfo {author} {\bibfnamefont
  {D.}~\bibnamefont {{Wands}}},\ }\bibfield  {title} {\enquote {\bibinfo
  {title} {{Density perturbations and black hole formation in hybrid
  inflation}},}\ }\href {https://doi.org/10.1103/PhysRevD.54.6040} {\bibfield
  {journal} {\bibinfo  {journal} {\prd}\ }\textbf {\bibinfo {volume} {54}},\
  \bibinfo {pages} {6040} (\bibinfo {year} {1996})},\ \Eprint
  {http://arxiv.org/abs/astro-ph/9605094} {astro-ph/9605094} \BibitemShut
  {NoStop}%
\bibitem [{\citenamefont {{Clesse}}\ and\ \citenamefont
  {{Garc{\'{\i}}a-Bellido}}(2015)}]{clesse+garcia-bellido_15}%
  \BibitemOpen
  \bibfield  {author} {\bibinfo {author} {\bibfnamefont {S.}~\bibnamefont
  {{Clesse}}}\ and\ \bibinfo {author} {\bibfnamefont {J.}~\bibnamefont
  {{Garc{\'{\i}}a-Bellido}}},\ }\bibfield  {title} {\enquote {\bibinfo {title}
  {Massive primordial black holes from hybrid inflation as dark matter and the
  seeds of galaxies},}\ }\href {https://doi.org/10.1103/PhysRevD.92.023524}
  {\bibfield  {journal} {\bibinfo  {journal} {\prd}\ }\textbf {\bibinfo
  {volume} {92}},\ \bibinfo {eid} {023524} (\bibinfo {year} {2015})},\ \Eprint
  {http://arxiv.org/abs/1501.07565} {arXiv:1501.07565} \BibitemShut {NoStop}%
\bibitem [{\citenamefont {{Garc{\'{\i}}a-Bellido}}\ and\ \citenamefont {{Ruiz
  Morales}}(2017)}]{garcia-bellido+morales_17}%
  \BibitemOpen
  \bibfield  {author} {\bibinfo {author} {\bibfnamefont {J.}~\bibnamefont
  {{Garc{\'{\i}}a-Bellido}}}\ and\ \bibinfo {author} {\bibfnamefont
  {E.}~\bibnamefont {{Ruiz Morales}}},\ }\bibfield  {title} {\enquote {\bibinfo
  {title} {{Primordial black holes from single field models of inflation}},}\
  }\href {https://doi.org/10.1016/j.dark.2017.09.007} {\bibfield  {journal}
  {\bibinfo  {journal} {Phys. Dark Univ.}\ }\textbf {\bibinfo {volume} {18}},\
  \bibinfo {pages} {47} (\bibinfo {year} {2017})},\ \Eprint
  {http://arxiv.org/abs/1702.03901} {arXiv:1702.03901} \BibitemShut {NoStop}%
\bibitem [{\citenamefont {{Hertzberg}}\ and\ \citenamefont
  {{Yamada}}(2018)}]{hertzberg+yamada_18}%
  \BibitemOpen
  \bibfield  {author} {\bibinfo {author} {\bibfnamefont {M.~P.}\ \bibnamefont
  {{Hertzberg}}}\ and\ \bibinfo {author} {\bibfnamefont {M.}~\bibnamefont
  {{Yamada}}},\ }\bibfield  {title} {\enquote {\bibinfo {title} {Primordial
  black holes from polynomial potentials in single field inflation},}\ }\href
  {https://doi.org/10.1103/PhysRevD.97.083509} {\bibfield  {journal} {\bibinfo
  {journal} {\prd}\ }\textbf {\bibinfo {volume} {97}},\ \bibinfo {eid} {083509}
  (\bibinfo {year} {2018})},\ \Eprint {http://arxiv.org/abs/1712.09750}
  {arXiv:1712.09750} \BibitemShut {NoStop}%
\bibitem [{\citenamefont {{Abbott}}\ \emph {et~al.}(2016)\citenamefont
  {{Abbott}} \emph {et~al.}}]{abbott+al_16}%
  \BibitemOpen
  \bibfield  {author} {\bibinfo {author} {\bibfnamefont {B.~P.}\ \bibnamefont
  {{Abbott}}} \emph {et~al.},\ }\bibfield  {title} {\enquote {\bibinfo {title}
  {Observation of gravitational waves from a binary black hole merger},}\
  }\href {https://doi.org/10.1103/PhysRevLett.116.061102} {\bibfield  {journal}
  {\bibinfo  {journal} {Phys. Rev. Lett.}\ }\textbf {\bibinfo {volume} {116}},\
  \bibinfo {eid} {061102} (\bibinfo {year} {2016})},\ \Eprint
  {http://arxiv.org/abs/1602.03837} {arXiv:1602.03837} \BibitemShut {NoStop}%
\bibitem [{\citenamefont {{Belczynski}}\ \emph {et~al.}(2010)\citenamefont
  {{Belczynski}}, \citenamefont {{Bulik}}, \citenamefont {{Fryer}},
  \citenamefont {{Ruiter}}, \citenamefont {{Valsecchi}}, \citenamefont
  {{Vink}},\ and\ \citenamefont {{Hurley}}}]{belczynski+al_10}%
  \BibitemOpen
  \bibfield  {author} {\bibinfo {author} {\bibfnamefont {K.}~\bibnamefont
  {{Belczynski}}}, \bibinfo {author} {\bibfnamefont {T.}~\bibnamefont
  {{Bulik}}}, \bibinfo {author} {\bibfnamefont {C.~L.}\ \bibnamefont
  {{Fryer}}}, \bibinfo {author} {\bibfnamefont {A.}~\bibnamefont {{Ruiter}}},
  \bibinfo {author} {\bibfnamefont {F.}~\bibnamefont {{Valsecchi}}}, \bibinfo
  {author} {\bibfnamefont {J.~S.}\ \bibnamefont {{Vink}}}, \ and\ \bibinfo
  {author} {\bibfnamefont {J.~R.}\ \bibnamefont {{Hurley}}},\ }\bibfield
  {title} {\enquote {\bibinfo {title} {On the maximum mass of stellar black
  holes},}\ }\href {https://doi.org/10.1088/0004-637X/714/2/1217} {\bibfield
  {journal} {\bibinfo  {journal} {Astrophy. J.}\ }\textbf {\bibinfo {volume}
  {714}},\ \bibinfo {pages} {1217} (\bibinfo {year} {2010})},\ \Eprint
  {http://arxiv.org/abs/0904.2784} {arXiv:0904.2784} \BibitemShut {NoStop}%
\bibitem [{\citenamefont {{Belczynski}}\ \emph {et~al.}(2016)\citenamefont
  {{Belczynski}}, \citenamefont {{Holz}}, \citenamefont {{Bulik}},\ and\
  \citenamefont {{O'Shaughnessy}}}]{belczynski+al_16}%
  \BibitemOpen
  \bibfield  {author} {\bibinfo {author} {\bibfnamefont {K.}~\bibnamefont
  {{Belczynski}}}, \bibinfo {author} {\bibfnamefont {D.~E.}\ \bibnamefont
  {{Holz}}}, \bibinfo {author} {\bibfnamefont {T.}~\bibnamefont {{Bulik}}}, \
  and\ \bibinfo {author} {\bibfnamefont {R.}~\bibnamefont {{O'Shaughnessy}}},\
  }\bibfield  {title} {\enquote {\bibinfo {title} {{The first
  gravitational-wave source from the isolated evolution of two stars in the
  40--100 solar mass range}},}\ }\href {https://doi.org/10.1038/nature18322}
  {\bibfield  {journal} {\bibinfo  {journal} {\nat}\ }\textbf {\bibinfo
  {volume} {534}},\ \bibinfo {pages} {512} (\bibinfo {year} {2016})},\ \Eprint
  {http://arxiv.org/abs/1602.04531} {arXiv:1602.04531} \BibitemShut {NoStop}%
\bibitem [{\citenamefont {{Rubin}}\ \emph {et~al.}(2001)\citenamefont
  {{Rubin}}, \citenamefont {{Sakharov}},\ and\ \citenamefont
  {{Khlopov}}}]{rubin+al_01}%
  \BibitemOpen
  \bibfield  {author} {\bibinfo {author} {\bibfnamefont {S.~G.}\ \bibnamefont
  {{Rubin}}}, \bibinfo {author} {\bibfnamefont {A.~S.}\ \bibnamefont
  {{Sakharov}}}, \ and\ \bibinfo {author} {\bibfnamefont {M.~Yu.}\ \bibnamefont
  {{Khlopov}}},\ }\bibfield  {title} {\enquote {\bibinfo {title} {The formation
  of primary galactic nuclei during phase transitions in the early universe},}\
  }\href {https://doi.org/10.1134/1.1385631} {\bibfield  {journal} {\bibinfo
  {journal} {J. of Exp. Theor. Phys.}\ }\textbf {\bibinfo {volume} {92}},\
  \bibinfo {pages} {921} (\bibinfo {year} {2001})},\ \Eprint
  {http://arxiv.org/abs/hep-ph/0106187} {hep-ph/0106187} \BibitemShut {NoStop}%
\bibitem [{\citenamefont {{Bean}}\ and\ \citenamefont
  {{Magueijo}}(2002)}]{bean+magueijo_02}%
  \BibitemOpen
  \bibfield  {author} {\bibinfo {author} {\bibfnamefont {R.}~\bibnamefont
  {{Bean}}}\ and\ \bibinfo {author} {\bibfnamefont {J.}~\bibnamefont
  {{Magueijo}}},\ }\bibfield  {title} {\enquote {\bibinfo {title} {Could
  supermassive black holes be quintessential primordial black holes?}}\ }\href
  {https://doi.org/10.1103/PhysRevD.66.063505} {\bibfield  {journal} {\bibinfo
  {journal} {\prd}\ }\textbf {\bibinfo {volume} {66}},\ \bibinfo {eid} {063505}
  (\bibinfo {year} {2002})},\ \Eprint {http://arxiv.org/abs/astro-ph/0204486}
  {astro-ph/0204486} \BibitemShut {NoStop}%
\bibitem [{\citenamefont {{D{\"u}chting}}(2004)}]{duechting_04}%
  \BibitemOpen
  \bibfield  {author} {\bibinfo {author} {\bibfnamefont {N.}~\bibnamefont
  {{D{\"u}chting}}},\ }\bibfield  {title} {\enquote {\bibinfo {title}
  {{Supermassive black holes from primordial black hole seeds}},}\ }\href
  {https://doi.org/10.1103/PhysRevD.70.064015} {\bibfield  {journal} {\bibinfo
  {journal} {\prd}\ }\textbf {\bibinfo {volume} {70}},\ \bibinfo {eid} {064015}
  (\bibinfo {year} {2004})},\ \Eprint {http://arxiv.org/abs/astro-ph/0406260}
  {astro-ph/0406260} \BibitemShut {NoStop}%
\bibitem [{\citenamefont {{Carr}}\ and\ \citenamefont
  {{Silk}}(2018)}]{carr+silk_18}%
  \BibitemOpen
  \bibfield  {author} {\bibinfo {author} {\bibfnamefont {B.}~\bibnamefont
  {{Carr}}}\ and\ \bibinfo {author} {\bibfnamefont {J.}~\bibnamefont
  {{Silk}}},\ }\bibfield  {title} {\enquote {\bibinfo {title} {Primordial black
  holes as generators of cosmic structures},}\ }\href
  {https://doi.org/10.1093/mnras/sty1204} {\bibfield  {journal} {\bibinfo
  {journal} {Mon. Not. R. Astro. Soc.}\ }\textbf {\bibinfo {volume} {478}},\
  \bibinfo {pages} {3756} (\bibinfo {year} {2018})},\ \Eprint
  {http://arxiv.org/abs/1801.00672} {arXiv:1801.00672} \BibitemShut {NoStop}%
\bibitem [{\citenamefont {{Chapline}}(1975)}]{chapline_75}%
  \BibitemOpen
  \bibfield  {author} {\bibinfo {author} {\bibfnamefont {G.~F.}\ \bibnamefont
  {{Chapline}}},\ }\bibfield  {title} {\enquote {\bibinfo {title} {Cosmological
  effects of primordial black holes},}\ }\href
  {https://doi.org/10.1038/253251a0} {\bibfield  {journal} {\bibinfo  {journal}
  {\nat}\ }\textbf {\bibinfo {volume} {253}},\ \bibinfo {pages} {251} (\bibinfo
  {year} {1975})}\BibitemShut {NoStop}%
\bibitem [{\citenamefont {{Bird}}\ \emph {et~al.}(2016)\citenamefont {{Bird}},
  \citenamefont {{Cholis}}, \citenamefont {{Mu{\~n}oz}}, \citenamefont
  {{Ali-Ha{\"i}moud}}, \citenamefont {{Kamionkowski}}, \citenamefont
  {{Kovetz}}, \citenamefont {{Raccanelli}},\ and\ \citenamefont
  {{Riess}}}]{bird+al_16}%
  \BibitemOpen
  \bibfield  {author} {\bibinfo {author} {\bibfnamefont {S.}~\bibnamefont
  {{Bird}}}, \bibinfo {author} {\bibfnamefont {I.}~\bibnamefont {{Cholis}}},
  \bibinfo {author} {\bibfnamefont {J.~B.}\ \bibnamefont {{Mu{\~n}oz}}},
  \bibinfo {author} {\bibfnamefont {Y.}~\bibnamefont {{Ali-Ha{\"i}moud}}},
  \bibinfo {author} {\bibfnamefont {M.}~\bibnamefont {{Kamionkowski}}},
  \bibinfo {author} {\bibfnamefont {E.~D.}\ \bibnamefont {{Kovetz}}}, \bibinfo
  {author} {\bibfnamefont {A.}~\bibnamefont {{Raccanelli}}}, \ and\ \bibinfo
  {author} {\bibfnamefont {A.~G.}\ \bibnamefont {{Riess}}},\ }\bibfield
  {title} {\enquote {\bibinfo {title} {Did {LIGO} detect dark matter?}}\ }\href
  {https://doi.org/10.1103/PhysRevLett.116.201301} {\bibfield  {journal}
  {\bibinfo  {journal} {Phys. Rev. Lett.}\ }\textbf {\bibinfo {volume} {116}},\
  \bibinfo {eid} {201301} (\bibinfo {year} {2016})},\ \Eprint
  {http://arxiv.org/abs/1603.00464} {arXiv:1603.00464} \BibitemShut {NoStop}%
\bibitem [{\citenamefont {{Carr}}\ \emph {et~al.}(2016)\citenamefont {{Carr}},
  \citenamefont {{K{\"u}hnel}},\ and\ \citenamefont {{Sandstad}}}]{carr+al_16}%
  \BibitemOpen
  \bibfield  {author} {\bibinfo {author} {\bibfnamefont {B.}~\bibnamefont
  {{Carr}}}, \bibinfo {author} {\bibfnamefont {F.}~\bibnamefont
  {{K{\"u}hnel}}}, \ and\ \bibinfo {author} {\bibfnamefont {M.}~\bibnamefont
  {{Sandstad}}},\ }\bibfield  {title} {\enquote {\bibinfo {title} {Primordial
  black holes as dark matter},}\ }\href
  {https://doi.org/10.1103/PhysRevD.94.083504} {\bibfield  {journal} {\bibinfo
  {journal} {\prd}\ }\textbf {\bibinfo {volume} {94}},\ \bibinfo {eid} {083504}
  (\bibinfo {year} {2016})},\ \Eprint {http://arxiv.org/abs/1607.06077}
  {arXiv:1607.06077} \BibitemShut {NoStop}%
\bibitem [{\citenamefont {{Dolgov}}\ and\ \citenamefont
  {{Silk}}(1993)}]{dolgov+silk_93}%
  \BibitemOpen
  \bibfield  {author} {\bibinfo {author} {\bibfnamefont {A.}~\bibnamefont
  {{Dolgov}}}\ and\ \bibinfo {author} {\bibfnamefont {J.}~\bibnamefont
  {{Silk}}},\ }\bibfield  {title} {\enquote {\bibinfo {title} {Baryon
  isocurvature fluctuations at small scales and baryonic dark matter},}\ }\href
  {https://doi.org/10.1103/PhysRevD.47.4244} {\bibfield  {journal} {\bibinfo
  {journal} {\prd}\ }\textbf {\bibinfo {volume} {47}},\ \bibinfo {pages} {4244}
  (\bibinfo {year} {1993})}\BibitemShut {NoStop}%
\bibitem [{\citenamefont {{Green}}(2016)}]{green_16}%
  \BibitemOpen
  \bibfield  {author} {\bibinfo {author} {\bibfnamefont {A.~M.}\ \bibnamefont
  {{Green}}},\ }\bibfield  {title} {\enquote {\bibinfo {title} {Microlensing
  and dynamical constraints on primordial black hole dark matter with an
  extended mass function},}\ }\href
  {https://doi.org/10.1103/PhysRevD.94.063530} {\bibfield  {journal} {\bibinfo
  {journal} {\prd}\ }\textbf {\bibinfo {volume} {94}},\ \bibinfo {eid} {063530}
  (\bibinfo {year} {2016})},\ \Eprint {http://arxiv.org/abs/1609.01143}
  {arXiv:1609.01143} \BibitemShut {NoStop}%
\bibitem [{\citenamefont {{Blinnikov}}\ \emph {et~al.}(2016)\citenamefont
  {{Blinnikov}}, \citenamefont {{Dolgov}}, \citenamefont {{Porayko}},\ and\
  \citenamefont {{Postnov}}}]{blinnikov+al_16}%
  \BibitemOpen
  \bibfield  {author} {\bibinfo {author} {\bibfnamefont {S.}~\bibnamefont
  {{Blinnikov}}}, \bibinfo {author} {\bibfnamefont {A.}~\bibnamefont
  {{Dolgov}}}, \bibinfo {author} {\bibfnamefont {N.~K.}\ \bibnamefont
  {{Porayko}}}, \ and\ \bibinfo {author} {\bibfnamefont {K.}~\bibnamefont
  {{Postnov}}},\ }\bibfield  {title} {\enquote {\bibinfo {title} {{Solving
  puzzles of GW150914 by primordial black holes}},}\ }\href
  {https://doi.org/10.1088/1475-7516/2016/11/036} {\bibfield  {journal}
  {\bibinfo  {journal} {J. Cosmol. Astropart. Phys.}\ }\textbf {\bibinfo
  {volume} {11}},\ \bibinfo {eid} {036} (\bibinfo {year} {2016})},\ \Eprint
  {http://arxiv.org/abs/1611.00541} {arXiv:1611.00541} \BibitemShut {NoStop}%
\bibitem [{\citenamefont {{Kannike}}\ \emph {et~al.}(2017)\citenamefont
  {{Kannike}}, \citenamefont {{Marzola}}, \citenamefont {{Raidal}},\ and\
  \citenamefont {{Veerm{\"a}e}}}]{kannike+al_17}%
  \BibitemOpen
  \bibfield  {author} {\bibinfo {author} {\bibfnamefont {K.}~\bibnamefont
  {{Kannike}}}, \bibinfo {author} {\bibfnamefont {L.}~\bibnamefont
  {{Marzola}}}, \bibinfo {author} {\bibfnamefont {M.}~\bibnamefont {{Raidal}}},
  \ and\ \bibinfo {author} {\bibfnamefont {H.}~\bibnamefont {{Veerm{\"a}e}}},\
  }\bibfield  {title} {\enquote {\bibinfo {title} {{Single field double
  inflation and primordial black holes}},}\ }\href
  {https://doi.org/10.1088/1475-7516/2017/09/020} {\bibfield  {journal}
  {\bibinfo  {journal} {J. Cosmol. Astropart. Phys.}\ }\textbf {\bibinfo
  {volume} {9}},\ \bibinfo {eid} {020} (\bibinfo {year} {2017})},\ \Eprint
  {http://arxiv.org/abs/1705.06225} {arXiv:1705.06225} \BibitemShut {NoStop}%
\bibitem [{\citenamefont {{K{\"u}hnel}}\ and\ \citenamefont
  {{Freese}}(2017)}]{kuhnel+freese_17}%
  \BibitemOpen
  \bibfield  {author} {\bibinfo {author} {\bibfnamefont {F.}~\bibnamefont
  {{K{\"u}hnel}}}\ and\ \bibinfo {author} {\bibfnamefont {K.}~\bibnamefont
  {{Freese}}},\ }\bibfield  {title} {\enquote {\bibinfo {title} {Constraints on
  primordial black holes with extended mass functions},}\ }\href
  {https://doi.org/10.1103/PhysRevD.95.083508} {\bibfield  {journal} {\bibinfo
  {journal} {\prd}\ }\textbf {\bibinfo {volume} {95}},\ \bibinfo {eid} {083508}
  (\bibinfo {year} {2017})},\ \Eprint {http://arxiv.org/abs/1701.07223}
  {arXiv:1701.07223} \BibitemShut {NoStop}%
\bibitem [{\citenamefont {{Carr}}\ \emph {et~al.}(2017)\citenamefont {{Carr}},
  \citenamefont {{Raidal}}, \citenamefont {{Tenkanen}}, \citenamefont
  {{Vaskonen}},\ and\ \citenamefont {{Veerm{\"a}e}}}]{carr+al_17}%
  \BibitemOpen
  \bibfield  {author} {\bibinfo {author} {\bibfnamefont {B.}~\bibnamefont
  {{Carr}}}, \bibinfo {author} {\bibfnamefont {M.}~\bibnamefont {{Raidal}}},
  \bibinfo {author} {\bibfnamefont {T.}~\bibnamefont {{Tenkanen}}}, \bibinfo
  {author} {\bibfnamefont {V.}~\bibnamefont {{Vaskonen}}}, \ and\ \bibinfo
  {author} {\bibfnamefont {H.}~\bibnamefont {{Veerm{\"a}e}}},\ }\bibfield
  {title} {\enquote {\bibinfo {title} {Primordial black hole constraints for
  extended mass functions},}\ }\href
  {https://doi.org/10.1103/PhysRevD.96.023514} {\bibfield  {journal} {\bibinfo
  {journal} {\prd}\ }\textbf {\bibinfo {volume} {96}},\ \bibinfo {eid} {023514}
  (\bibinfo {year} {2017})},\ \Eprint {http://arxiv.org/abs/1705.05567}
  {arXiv:1705.05567} \BibitemShut {NoStop}%
\bibitem [{\citenamefont {{Josan}}\ \emph {et~al.}(2009)\citenamefont
  {{Josan}}, \citenamefont {{Green}},\ and\ \citenamefont
  {{Malik}}}]{josan+al_09}%
  \BibitemOpen
  \bibfield  {author} {\bibinfo {author} {\bibfnamefont {A.~S.}\ \bibnamefont
  {{Josan}}}, \bibinfo {author} {\bibfnamefont {A.~M.}\ \bibnamefont
  {{Green}}}, \ and\ \bibinfo {author} {\bibfnamefont {K.~A.}\ \bibnamefont
  {{Malik}}},\ }\bibfield  {title} {\enquote {\bibinfo {title} {Generalized
  constraints on the curvature perturbation from primordial black holes},}\
  }\href {https://doi.org/10.1103/PhysRevD.79.103520} {\bibfield  {journal}
  {\bibinfo  {journal} {\prd}\ }\textbf {\bibinfo {volume} {79}},\ \bibinfo
  {eid} {103520} (\bibinfo {year} {2009})},\ \Eprint
  {http://arxiv.org/abs/0903.3184} {arXiv:0903.3184} \BibitemShut {NoStop}%
\bibitem [{\citenamefont {{Carr}}\ \emph {et~al.}(2010)\citenamefont {{Carr}},
  \citenamefont {{Kohri}}, \citenamefont {{Sendouda}},\ and\ \citenamefont
  {{Yokoyama}}}]{carr+al_10}%
  \BibitemOpen
  \bibfield  {author} {\bibinfo {author} {\bibfnamefont {B.~J.}\ \bibnamefont
  {{Carr}}}, \bibinfo {author} {\bibfnamefont {K.}~\bibnamefont {{Kohri}}},
  \bibinfo {author} {\bibfnamefont {Y.}~\bibnamefont {{Sendouda}}}, \ and\
  \bibinfo {author} {\bibfnamefont {J.}~\bibnamefont {{Yokoyama}}},\ }\bibfield
   {title} {\enquote {\bibinfo {title} {{New cosmological constraints on
  primordial black holes}},}\ }\href
  {https://doi.org/10.1103/PhysRevD.81.104019} {\bibfield  {journal} {\bibinfo
  {journal} {\prd}\ }\textbf {\bibinfo {volume} {81}},\ \bibinfo {eid} {104019}
  (\bibinfo {year} {2010})},\ \Eprint {http://arxiv.org/abs/0912.5297}
  {arXiv:0912.5297} \BibitemShut {NoStop}%
\bibitem [{\citenamefont {{Inomata}}\ \emph {et~al.}(2018)\citenamefont
  {{Inomata}}, \citenamefont {{Kawasaki}}, \citenamefont {{Mukaida}},\ and\
  \citenamefont {{Yanagida}}}]{inomata+al_18}%
  \BibitemOpen
  \bibfield  {author} {\bibinfo {author} {\bibfnamefont {K.}~\bibnamefont
  {{Inomata}}}, \bibinfo {author} {\bibfnamefont {M.}~\bibnamefont
  {{Kawasaki}}}, \bibinfo {author} {\bibfnamefont {K.}~\bibnamefont
  {{Mukaida}}}, \ and\ \bibinfo {author} {\bibfnamefont {T.~T.}\ \bibnamefont
  {{Yanagida}}},\ }\bibfield  {title} {\enquote {\bibinfo {title} {Double
  inflation as a single origin of primordial black holes for all dark matter
  and {LIGO} observations},}\ }\href
  {https://doi.org/10.1103/PhysRevD.97.043514} {\bibfield  {journal} {\bibinfo
  {journal} {\prd}\ }\textbf {\bibinfo {volume} {97}},\ \bibinfo {eid} {043514}
  (\bibinfo {year} {2018})},\ \Eprint {http://arxiv.org/abs/1711.06129}
  {arXiv:1711.06129} \BibitemShut {NoStop}%
\bibitem [{\citenamefont {{Inomata}}\ \emph {et~al.}(2017)\citenamefont
  {{Inomata}}, \citenamefont {{Kawasaki}}, \citenamefont {{Mukaida}},
  \citenamefont {{Tada}},\ and\ \citenamefont {{Yanagida}}}]{inomata+al_17}%
  \BibitemOpen
  \bibfield  {author} {\bibinfo {author} {\bibfnamefont {K.}~\bibnamefont
  {{Inomata}}}, \bibinfo {author} {\bibfnamefont {M.}~\bibnamefont
  {{Kawasaki}}}, \bibinfo {author} {\bibfnamefont {K.}~\bibnamefont
  {{Mukaida}}}, \bibinfo {author} {\bibfnamefont {Y.}~\bibnamefont {{Tada}}}, \
  and\ \bibinfo {author} {\bibfnamefont {T.~T.}\ \bibnamefont {{Yanagida}}},\
  }\bibfield  {title} {\enquote {\bibinfo {title} {{Inflationary primordial
  black holes as all dark matter}},}\ }\href
  {https://doi.org/10.1103/PhysRevD.96.043504} {\bibfield  {journal} {\bibinfo
  {journal} {\prd}\ }\textbf {\bibinfo {volume} {96}},\ \bibinfo {eid} {043504}
  (\bibinfo {year} {2017})},\ \Eprint {http://arxiv.org/abs/1701.02544}
  {arXiv:1701.02544} \BibitemShut {NoStop}%
\bibitem [{\citenamefont {{Press}}\ and\ \citenamefont
  {{Schechter}}(1974)}]{press+schechter_74}%
  \BibitemOpen
  \bibfield  {author} {\bibinfo {author} {\bibfnamefont {W.~H.}\ \bibnamefont
  {{Press}}}\ and\ \bibinfo {author} {\bibfnamefont {P.}~\bibnamefont
  {{Schechter}}},\ }\bibfield  {title} {\enquote {\bibinfo {title} {Formation
  of galaxies and clusters of galaxies by self-similar gravitational
  condensation},}\ }\href {https://doi.org/10.1086/152650} {\bibfield
  {journal} {\bibinfo  {journal} {\apj}\ }\textbf {\bibinfo {volume} {187}},\
  \bibinfo {pages} {425} (\bibinfo {year} {1974})}\BibitemShut {NoStop}%
\bibitem [{\citenamefont {{Hartle}}\ and\ \citenamefont
  {{Hertog}}(2013)}]{hartle+hertog_13}%
  \BibitemOpen
  \bibfield  {author} {\bibinfo {author} {\bibfnamefont {J.}~\bibnamefont
  {{Hartle}}}\ and\ \bibinfo {author} {\bibfnamefont {T.}~\bibnamefont
  {{Hertog}}},\ }\bibfield  {title} {\enquote {\bibinfo {title} {Anthropic
  bounds on {$\Lambda$} from the no-boundary quantum state},}\ }\href
  {https://doi.org/10.1103/PhysRevD.88.123516} {\bibfield  {journal} {\bibinfo
  {journal} {\prd}\ }\textbf {\bibinfo {volume} {88}},\ \bibinfo {eid} {123516}
  (\bibinfo {year} {2013})},\ \Eprint {http://arxiv.org/abs/1309.0493}
  {arXiv:1309.0493} \BibitemShut {NoStop}%
\bibitem [{\citenamefont {{Ade}}\ \emph {et~al.}(2014)\citenamefont {{Ade}}
  \emph {et~al.}}]{planck_13INF}%
  \BibitemOpen
  \bibfield  {author} {\bibinfo {author} {\bibfnamefont {P.~A.~R.}\
  \bibnamefont {{Ade}}} \emph {et~al.},\ }\bibfield  {title} {\enquote
  {\bibinfo {title} {{{\it Planck} 2013 results. XXII. Constraints on
  inflation}},}\ }\href {https://doi.org/10.1051/0004-6361/201321569}
  {\bibfield  {journal} {\bibinfo  {journal} {Astron. Astrophys.}\ }\textbf
  {\bibinfo {volume} {571}},\ \bibinfo {eid} {A22} (\bibinfo {year} {2014})},\
  \Eprint {http://arxiv.org/abs/1303.5082} {arXiv:1303.5082} \BibitemShut
  {NoStop}%
\bibitem [{\citenamefont {{Ade}}\ \emph
  {et~al.}(2016{\natexlab{a}})\citenamefont {{Ade}} \emph
  {et~al.}}]{ade+al_15}%
  \BibitemOpen
  \bibfield  {author} {\bibinfo {author} {\bibfnamefont {P.~A.~R.}\
  \bibnamefont {{Ade}}} \emph {et~al.},\ }\bibfield  {title} {\enquote
  {\bibinfo {title} {{{\it Planck} 2015 results. XX. Constraints on
  inflation}},}\ }\href {https://doi.org/10.1051/0004-6361/201525898}
  {\bibfield  {journal} {\bibinfo  {journal} {Astron. Astrophys.}\ }\textbf
  {\bibinfo {volume} {594}},\ \bibinfo {eid} {A20} (\bibinfo {year}
  {2016}{\natexlab{a}})},\ \Eprint {http://arxiv.org/abs/1502.02114}
  {arXiv:1502.02114} \BibitemShut {NoStop}%
\bibitem [{\citenamefont {{Ade}}\ \emph
  {et~al.}(2016{\natexlab{b}})\citenamefont {{Ade}} \emph
  {et~al.}}]{bicep2+keck_16}%
  \BibitemOpen
  \bibfield  {author} {\bibinfo {author} {\bibfnamefont {P.~A.~R.}\
  \bibnamefont {{Ade}}} \emph {et~al.},\ }\bibfield  {title} {\enquote
  {\bibinfo {title} {Improved constraints on cosmology and foregrounds from
  {BICEP2 and Keck Array} cosmic microwave background data with inclusion of 95
  {GHz} band},}\ }\href {https://doi.org/10.1103/PhysRevLett.116.031302}
  {\bibfield  {journal} {\bibinfo  {journal} {Phys. Rev. Lett.}\ }\textbf
  {\bibinfo {volume} {116}},\ \bibinfo {eid} {031302} (\bibinfo {year}
  {2016}{\natexlab{b}})},\ \Eprint {http://arxiv.org/abs/1510.09217}
  {arXiv:1510.09217} \BibitemShut {NoStop}%
\bibitem [{\citenamefont {{Ade}}\ \emph
  {et~al.}(2016{\natexlab{c}})\citenamefont {{Ade}} \emph
  {et~al.}}]{planck_15_CosmologicalParameters}%
  \BibitemOpen
  \bibfield  {author} {\bibinfo {author} {\bibfnamefont {P.~A.~R.}\
  \bibnamefont {{Ade}}} \emph {et~al.},\ }\bibfield  {title} {\enquote
  {\bibinfo {title} {{{\it Planck} 2015 results. XIII. Cosmological
  parameters}},}\ }\href {https://doi.org/10.1051/0004-6361/201525830}
  {\bibfield  {journal} {\bibinfo  {journal} {Astron. Astrophys.}\ }\textbf
  {\bibinfo {volume} {594}},\ \bibinfo {eid} {A13} (\bibinfo {year}
  {2016}{\natexlab{c}})},\ \Eprint {http://arxiv.org/abs/1502.01589}
  {arXiv:1502.01589} \BibitemShut {NoStop}%
\bibitem [{\citenamefont {Weinberg}(2015)}]{weinberg_15}%
  \BibitemOpen
  \bibfield  {author} {\bibinfo {author} {\bibfnamefont {S.}~\bibnamefont
  {Weinberg}},\ }\href@noop {} {\emph {\bibinfo {title} {To Explain the World:
  The Discovery of Modern Science}}}\ (\bibinfo  {publisher} {Harper, New
  York},\ \bibinfo {year} {2015})\BibitemShut {NoStop}%
\bibitem [{\citenamefont {Lewens}(2007)}]{lewens_07}%
  \BibitemOpen
  \bibfield  {author} {\bibinfo {author} {\bibfnamefont {T.}~\bibnamefont
  {Lewens}},\ }\href@noop {} {\emph {\bibinfo {title} {Darwin}}}\ (\bibinfo
  {publisher} {Routledge, London},\ \bibinfo {year} {2007})\BibitemShut
  {NoStop}%
\bibitem [{\citenamefont {Bialek}(2012)}]{bialek_12}%
  \BibitemOpen
  \bibfield  {author} {\bibinfo {author} {\bibfnamefont {W.}~\bibnamefont
  {Bialek}},\ }\href@noop {} {\emph {\bibinfo {title} {Biophysics: {S}earching
  for {P}rinciples}}}\ (\bibinfo  {publisher} {Princeton University Press,
  Princeton},\ \bibinfo {year} {2012})\BibitemShut {NoStop}%
\bibitem [{\citenamefont {{Bialek}}(2018)}]{bialek_15}%
  \BibitemOpen
  \bibfield  {author} {\bibinfo {author} {\bibfnamefont {W.}~\bibnamefont
  {{Bialek}}},\ }\bibfield  {title} {\enquote {\bibinfo {title} {Perspectives
  on theory at the interface of physics and biology},}\ }\href
  {https://doi.org/10.1088/1361-6633/aa995b} {\bibfield  {journal} {\bibinfo
  {journal} {Rep. Prog. Phys.}\ }\textbf {\bibinfo {volume} {81}},\ \bibinfo
  {eid} {012601} (\bibinfo {year} {2018})},\ \Eprint
  {http://arxiv.org/abs/1512.08954} {arXiv:1512.08954} \BibitemShut {NoStop}%
\bibitem [{\citenamefont {Kuhn}(1977)}]{kuhn_77}%
  \BibitemOpen
  \bibfield  {author} {\bibinfo {author} {\bibfnamefont {T.~S.}\ \bibnamefont
  {Kuhn}},\ }\href@noop {} {\emph {\bibinfo {title} {The Essential Tension:
  Selected Studies in Scientific Tradition and Change}}}\ (\bibinfo
  {publisher} {University of Chicago Press, Chicago},\ \bibinfo {year} {1977})\
  Chap.~\bibinfo {chapter} {13}\BibitemShut {NoStop}%
\bibitem [{\citenamefont {McMullin}(1982)}]{mcmullin_83}%
  \BibitemOpen
  \bibfield  {author} {\bibinfo {author} {\bibfnamefont {E.}~\bibnamefont
  {McMullin}},\ }\bibfield  {title} {\enquote {\bibinfo {title} {Values in
  science},}\ }\href {http://www.jstor.org/stable/192409} {\bibfield  {journal}
  {\bibinfo  {journal} {PSA: Proceedings of the Biennial Meeting of the
  Philosophy of Science Association}\ }\textbf {\bibinfo {volume} {1982}},\
  \bibinfo {pages} {3} (\bibinfo {year} {1982})}\BibitemShut {NoStop}%
\end{thebibliography}%

\end{document}